\documentclass{ametsocV6.1}

\usepackage{subcaption}

\usepackage{booktabs}

\title{Ecohydrological Controls on Moist Convection\\ and Long-Term Rainfall Feedback}
\authors{
Elizabeth Cultra,\aff{a}
Jun Yin,\aff{b}
Mark Bartlett,\aff{c} and
\correspondingauthor{Amilcare Porporato, aporpora@princeton.edu}{Amilcare Porporato}\aff{a,d}
}

\affiliation{\aff{a}{Department of Civil and Environmental Engineering, Princeton University, Princeton, New Jersey, USA}\\
\aff{b}{School of Hydrology and Water Resources, Nanjing University of Information Science and Technology (NUIST), Nanjing, China} \\
\aff{c}{The Water Institute, Baton Rouge, Louisiana, USA} \\ 
\aff{d}{The High Meadows Environmental Institute, Princeton University, Princeton, New Jersey, USA}
}

\abstract{To elucidate how land surface state and soil moisture dynamics regulate moist convection, and how convective rainfall subsequently reshapes surface and root-zone hydrology, we develop a stochastic dynamical model that couples soil moisture, vegetation hydraulics, atmospheric boundary layer evolution, and convective available potential energy (CAPE). We show that CAPE depends not only on the free-tropospheric environment but also on soil moisture, through its control of surface fluxes, boundary-layer growth, and the timing of the intersection between the atmospheric boundary layer and the lifting condensation level (LCL). Soil texture and plant properties strongly modulate convective potential during dry-down. Loamy sand favors convection at relatively high soil moisture and maintains the largest CAPE at the time of LCL-ABL crossing across drying conditions. In contrast, sandy soils exhibit high CAPE when wet but lose convective potential rapidly as they dry. As matric potential becomes more negative, convection is increasingly suppressed in finer, loamy clay textures. Plant functional type further shapes dry-down dynamics: water-use-maximizing strategies can enhance dry persistence via stomatal closure during drying, whereas more conservative strategies can sustain convection for longer periods. On longer timescales, stochastic rainfall forcing with CAPE-dependent precipitation intensity produces persistent wet and dry soil moisture regimes, with switching times that depend on soil hydraulic properties, plant physiological traits, and atmospheric conditions.
}

\begin{document}

\maketitle
\section*{Significance Statement}

Identifying the key processes and timescales that balance land–atmosphere interactions is a central question in climate science. In this work, we utilize an ecohydrological model in which vegetation, soil type, and atmospheric properties jointly regulate convective rainfall. The feedback loop is maintained through dynamic surface fluxes, which modulate precipitation frequency and intensity. We find that this feedback may promote both wet and dry climate persistence, highlighting the significant role of the land–surface hydrologic state in influencing the atmosphere.

\section{Introduction}

The water and energy budgets are tightly coupled at the land–atmosphere interface. The land surface regulates moisture availability and surface energy fluxes, thereby shaping the state of the lower atmosphere. This influence is mediated through the growth of the atmospheric boundary layer (ABL), which dynamically integrates surface forcing with conditions in the free atmosphere, linking processes across multiple spatial and temporal scales. Moreover, hydrologic conditions impart a memory effect through soil moisture storage and dynamics. Understanding these multiscale processes is increasingly important as rising greenhouse gas emissions amplify lower-tropospheric water vapor and drive substantial changes in the global hydrologic cycle \citep{held2006robust}. Higher global temperatures intensify precipitation extremes, many of which occur as convective rainfall events \citep{berg2013strong}, and these events are strongly modulated by land-surface conditions \citep{trier2004study, seneviratne2010investigating, emanuel2023physics}.

Considerable work has focused on land–atmosphere interactions, particularly the role of land-surface fluxes and soil moisture dynamics in shaping ABL development and triggering convective precipitation. In this context, soil moisture tends to amplify the magnitude and frequency of convective events through a general positive feedback mechanism that increases moist static energy within the ABL \citep{pal2001pathways}. Previous research indicates that the strength of soil–precipitation coupling depends on the partitioning of surface fluxes, with the competing effects of sensible and latent heating altering the height of the ABL and lifting condensation level, contingent on initial atmospheric conditions \citep{findell2003atmospheric}.

Critical to these interactions is convective available potential energy (CAPE), whose buildup is strongly related to the manifestation of extreme rainfall \citep{emanuel1997overview}. Through CAPE and its dynamic evolution, the timing, initiation, and maintenance of cloud formation associated with moist convection are directly tied to surface fluxes of water vapor and energy in a highly nonlinear way. These surface fluxes depend on land cover and soil properties, which therefore exert strong control over climate drivers \citep{bonan2008forests}. Several land-surface models have parameterized vegetation fluxes for GCMs \citep{lawrence2019community}, whereas similar attempts have been more limited in fully coupled land–atmosphere frameworks \citep{chaney2016deriving, emanuel2023physics, waterman2025surface}. 

Regarding the long-term dynamics of land–atmosphere interactions, analyses of the two-way coupling between ecohydrological dynamics and atmospheric convection have largely been confined to either simplified stochastic minimalist frameworks \citep{d2004a} or complex numerical simulations \citep{cioni2017effect}. Both approaches, despite their utility, are limited in isolating fundamental ecohydrological controls on long-term hydrologic regimes. Early minimalist models linked the probability of precipitation occurrence to soil moisture states, highlighting the potential emergence of multiple preferential moisture regimes and their connection to the persistence of climate anomalies \citep{d2004a, d2004b}. In parallel, comprehensive atmospheric numerical studies employing both fine-scale resolved convection and coarse-scale parameterized convection have been used to understand the influence of surface fluxes and horizontal circulation on convective initiation. These numerical simulation efforts provide realism through explicit representation of physical processes, but they encounter substantial computational constraints, numerical sensitivity, and difficulty in representing land-surface heterogeneity and ecohydrological processes with sufficient realism and detail. In particular, the role of plant water-use strategies and the consequences of variation in hydraulic and photosynthetic traits on surface water and energy fluxes, and ultimately on ABL growth and CAPE, remain insufficiently investigated.

Our main goal in this work is to investigate the ecohydrological dimension of land–rainfall feedbacks, with emphasis placed on their multiscale temporal structure: from the hourly diurnal evolution of land–atmosphere exchange to the influence of soil and plant properties on moist convective potential and, ultimately, on long-term rainfall statistics. These interactions have not yet been assessed across a continuum of plant functional types within a dynamical soil moisture framework. While prior parameterized convection models suggest that increased vegetation cover promotes convective initiation when soil moisture is not limiting \citep{clark1995numerical}, here we introduce a hydraulically explicit approach that links plant physiological traits, soil type, and the long-term consequences of convective triggering for climate states. By resolving the soil–plant–atmosphere continuum in a thermodynamically consistent way, we reveal the soil–plant control on water potential gradients (normalized specific Gibbs free energy) as the thermodynamically consistent potential for water transport between the land and the atmosphere. This allows us to consistently constrain such fluxes based on soil and plant hydraulic properties and land-surface energetic constraints \citep{porporato2022ecohydrology}. Related work increasingly emphasizes the role of plant physiology and phenology in convective precipitation \citep{chapman2021soil}, particularly under warming trends and the associated shifts in water exchange between the troposphere and the atmospheric boundary layer. Daytime CAPE/CIN buildup and erosion have been shown to be highly sensitive to land–surface conductance, i.e., to soil and vegetation conditions \citep{emanuel2023physics}. These surface properties regulate the transition between water-limited and energy-limited regimes. Building on this mechanistic understanding and on \cite{yin2015land}, we directly link surface moisture and heat fluxes to characteristic variations in plant hydraulic traits and soil type, embedded within a physically grounded stochastic rainfall model. This framework enables long-term evaluation of feedbacks on the terrestrial water balance while explicitly representing uncertainty in unresolved land–atmosphere processes.

We focus specifically on the temporal dynamics of this feedback and the roles of soil and plant types, assuming homogeneous spatial conditions; the role of spatial heterogeneities, while important \citep{hohenegger2018role}, will be explored in future contributions. We therefore develop a process-based stochastic framework relating plant physiology, atmospheric boundary layer dynamics, and the initiation and intensity of convective rainfall, linking the plant-soil moisture model to ABL growth and CAPE evolution, including stomatal regulation and plant hydraulic vulnerability. 
We then explore the model over a period of no rainfall (i.e., a dry-down) to map the relationship between convective indicators and soil moisture, dependent on both plant and soil type. To proceed to longer timescales, we extend the work of \cite{muller2011intensification} by adding thermodynamic and mass constraints to convective precipitation totals, thereby relating atmospheric profiles to surface and atmospheric boundary layer evolution. We embed these land–atmosphere feedbacks in a stochastic rainfall representation, which—once propagated to soil moisture dynamics—can lead to bimodal and persistent states. 
This allows us to analyze how plants may actively and passively regulate properties of the atmosphere–soil moisture feedback on both diurnal and seasonal timescales.  

\section{Methodology}

The primary interaction between the land surface and the atmosphere occurs through the diurnal evolution of the atmospheric boundary layer (ABL), whose growth is primarily driven by sensible heat flux. In turn, this flux depends on the land-surface partitioning of incoming solar radiation into sensible and latent heat (Eq.~\eqref{eq: surfeng}), with the latter controlled by soil moisture dynamics and the associated ecohydrological processes within the soil–plant system. Assuming dry adiabatic lifting from the near-surface, the LCL height depends directly on the ABL temperature and humidity (see Eq.~\eqref{app: LCLh}), which in turn modify the level of free convection and thus CAPE, as detailed in Eq.~\eqref{eq: CAPE}.  Accordingly, to isolate the essential elements of such interaction, we couple the well-established well-mixed ABL equations, Eqs.~\eqref{eq: dhdt},~\eqref{eq: dqdt}, and~\eqref{eq: dthetadt}, \citep{stull1976energetics, Stull1988, garratt1994atmospheric} with a vertically averaged soil moisture balance over the rooting zone given as Eq.~\eqref{balance-eq} \citep{porporato2022ecohydrology}.

\subsection{Coupled ABL and Soil Moisture Dynamics}

For the diurnal growth of the ABL, we assume that buoyancy-generated turbulence dominates the entrainment flux, thereby closing the vertically integrated system. For a multi-day evolution, we consider a collapse of the boundary layer during rainfall events or at night. The height of the well-mixed ABL for clear-sky conditions \citep{stull1976energetics, Stull1988,garratt1994atmospheric, porporato2009atmospheric, porporato2022ecohydrology} is proportional to the virtual sensible heat flux, $H_v$:
\begin{equation}
\frac{dh}{dt} = \frac{(1+2\beta) H_v}{\rho_a c_p \gamma_{\theta_v} h},
\label{eq: dhdt}
\end{equation}
where $\gamma_{\theta_{v,f}}$ is the vertical slope of the virtual potential in the free atmosphere, $\beta\approx 0.2$ is the entrainment ratio, $c_p$ is the specific heat of dry air, and $\rho_a$ is the density of air. The virtual sensible heat flux is $H_v = H \ + 0.61\theta_{\text{BL}} c_p \rho_w\Theta$, where $H = \rho_a c_p g_a (\theta_l-\theta_{\text{BL}})$, including $\theta_l-\theta_{\text{BL}}$, the difference between the leaf surface and boundary layer potential temperature. Evapotranspiration is denoted as ($\Theta$), and $g_a$ is the atmospheric conductance. Virtual potential temperature ($\theta_v$) is used to account for changes in density from water vapor and liquid water \citep{smith2013physics}: 

\begin{equation}
\theta_{v} = T\left(\frac{p_0}{p}\right)^{\frac{R_d}{c_p}}\left[1+q\left(\frac{1}{\varphi}-1\right)\right] = \left(\frac{p_0}{p}\right)^{\frac{R_d}{c_p}} T_v, 
\label{eq: virtualtemp}
\end{equation}
where $\varphi = R_d/R_v$, the gas constants for water vapor and dry air, respectively.  Liquid water is considered zero in the cloud-free atmosphere, just prior to the onset of convection. The change in specific humidity ($q_{\text{BL}}$) within the boundary layer is a function of surface and entrainment fluxes and is directly coupled to the free-atmospheric environment above. The entrainment flux is equal to the derivative of the boundary-layer height multiplied by the difference between the boundary layer and free atmospheric virtual temperature at the capping inversion height, $h(t)$:
\begin{equation}
\rho_a h \frac{d q_{\text{BL}}}{d t} =\rho_w\Theta+\rho_a\left[q_{f}(h)-q_{\text{BL}}\right] \frac{d h}{d t},
\label{eq: dqdt}
\end{equation}
where the input of water vapor through $\Theta$, regulated by soil moisture, increases the boundary layer humidity, while entrainment of free atmospheric vapor at the boundary layer inversion, $q_f(h)$, is modified by the rate of change of the boundary layer height.

The change in ABL virtual potential temperature is similarly controlled by virtual sensible heat and entrainment from the free atmosphere:
\begin{equation}
\rho_a c_p h \frac{d \theta_{v, \text{BL}}}{d t}=H_v+\rho_a c_p\left[\theta_{v,f}(h)-\theta_{v, \text{BL}}\right] \frac{d h}{d t},
\label{eq: dthetadt}
\end{equation}
where $\theta_{v,f}(h)$ is the free atmospheric virtual potential temperature (see Appendix \ref{app: LCLh}) at the boundary layer height.

The partitioning of latent and sensible heat is given by the quasi-equilibrium surface energy balance, as 
\begin{equation}
\phi = L_v\Theta\ +  H +G,
\label{eq: surfeng}
\end{equation}
where $L_v$ is the latent heat of vaporization and $\phi$ is the net radiation to the surface. The term $G$, soil heat flux to the lower layers, may be neglected when modeling due to the low heat capacity of tree canopies. Here, the land conditions (i.e, the vegetation and soil) appear in the atmospheric dynamics, directly connected to both sensible and latent heat. Plant functional traits govern the partitioning of these fluxes, while soil moisture availability ultimately drives the timing of these physiological responses. These hydraulic strategies, in turn, control the boundary layer dynamics during the day and shape hydrologic partitioning over timescales.

To capture land-surface impacts and related feedbacks, we couple a soil moisture model \citep{porporato2022ecohydrology} with the previous ABL equations. This is obtained by vertically averaging the Richards equation over the root zone,   
\begin{equation}
nZ_r\frac{ds}{dt} = I_{T} -  \Theta - L,
\label{balance-eq}
\end{equation}
where $s$ is the relative soil moisture, $Z_r$ is the active rooting depth and $n$ is the porosity. The fluxes on the right side of Eq.~\eqref{balance-eq} are, respectively, infiltration into the soil driven by large-scale processes and convection, which includes precipitation and runoff (i.e., $I_T = P - Q$), while losses occur due to evapotranspiration ($\Theta$) and leakage ($L$). We assume well-vegetated conditions, so that transpiration ($\Theta$) dominates over bare-soil evaporation, which we do not consider here for simplicity. These terms are a function of relative soil moisture, $s\in (0,1]$, along with atmospheric parameters such as boundary layer temperature, humidity, and net available energy, and are thus directly coupled to the ABL dynamics. 

\begin{figure}[h]
        \centering
        \includegraphics[width=1\linewidth]{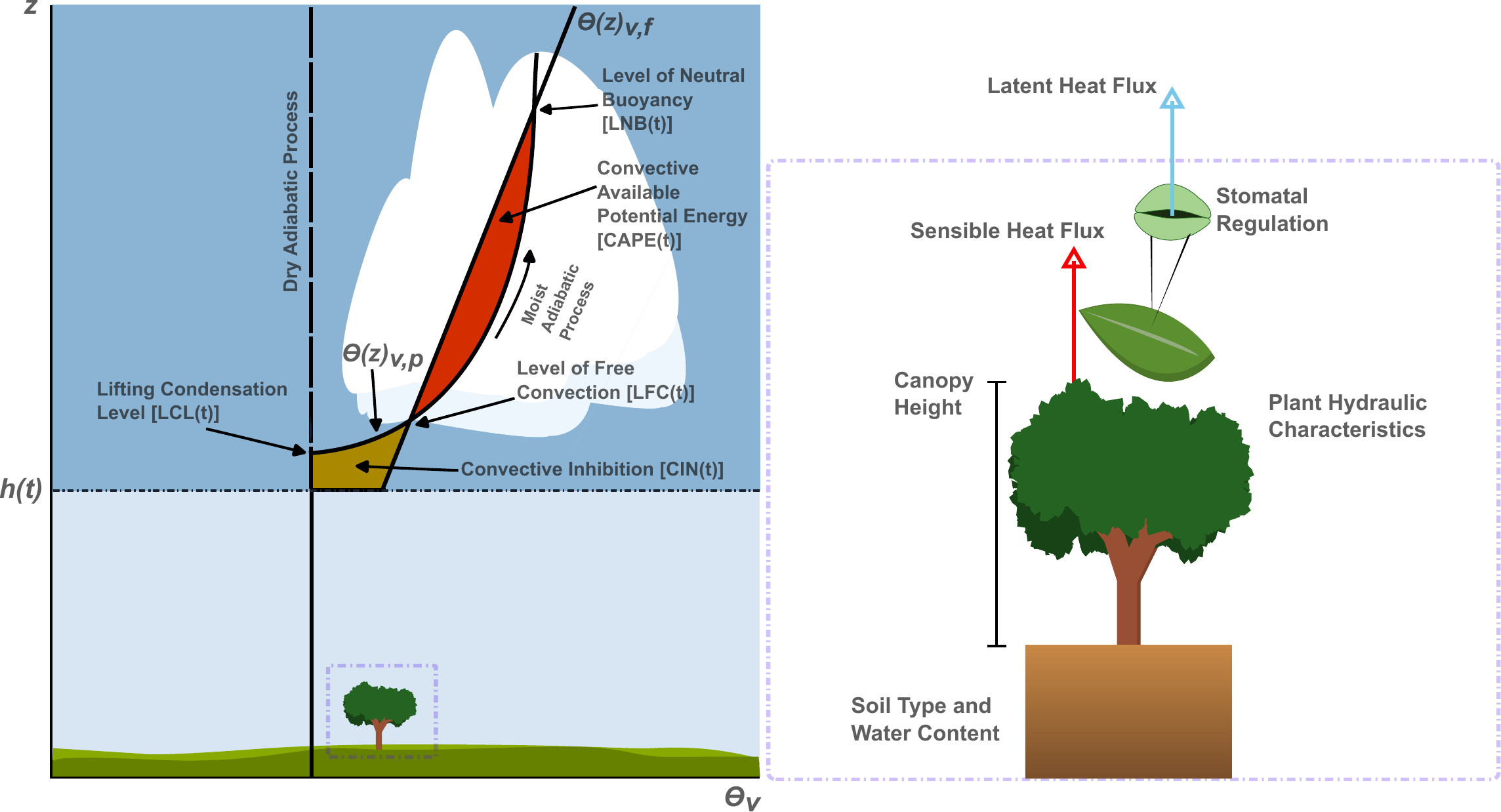}
        \caption{Schematic showing the temperature profiles within the atmospheric boundary layer and in the free atmosphere, including the moist temperature profile as in Eq.~\eqref{eq: moist lapse rate}. The shaded yellow and red areas between the two temperature profiles represent CIN and CAPE, respectively. It is assumed that a parcel follows a dry adiabatic process until it reaches the LCL.}
        \label{fig: diagram}
\end{figure}

These coupled soil moisture and ABL dynamics impact atmospheric convection potential through the time of ABL crossing of the LCL (see Figure \ref{fig: diagram}) as well as through the values of CAPE, 
\begin{equation}
\label{eq: CAPE}
\operatorname{CAPE}(t) =\int_{z_{\text{LFC}}(t)}^{z_{\text{LNB}} (t)} g \frac{T_{v, p}(z, t)-T_{v, f}(z, t)}{T_{v, f}(z,t)} d z,
\end{equation}
i.e., the integral of the difference between the surrounding free air virtual temperature, $T_{v,f}$, and the virtual temperature of the adiabatically raised air parcel following the saturated lapse rate, $T_{v, p}$, related to potential temperature given in Eq.~\eqref{eq: virtualtemp}. Physically, CAPE represents the area of positive buoyancy between the level of free convection (LFC) and the level of neutral buoyancy (LNB), where the lifted parcel is warmer and less dense than the free atmosphere, indicated by the red shaded region in Figure \ref{fig: diagram}.

Notably, the unit-mass acceleration due to parcel buoyancy is directly related to its maximum updraft speed, i.e., $\frac{w_{\text{max}}^2}{2} = \text{CAPE} $, and when precipitation occurs, also to rainfall intensity. In this way, soil moisture directly influences CAPE by regulating surface temperature and humidity, thereby affecting the LCL height, which in turn influences both parcel height and the integral bounds of CAPE. 

In the absence of considerable changes in plant water storage \citep{cultra2025plant}, the transpiration flux through the stomata and into the atmosphere  is equal to the liquid water flux through the plant from the soil to the leaf, given by the difference in water potential from the leaf to the soil, multiplied by the soil-root-plant conductance, 
\begin{equation}
\Theta =g_{st, a} \frac{\rho_a}{\rho_w}(q_l\left(\theta_l, \psi_l\right)-q_{\text{BL}}(\theta_{\text{BL}}))= g_{srp}(\psi_s(s) - \psi_l).
\label{equationtrans}
\end{equation}
The flux through the stomata into the atmosphere is given by the difference in specific humidity between the leaf's internal, $q_l(\theta_l, \psi_l)$, and the lower atmospheric environment, $q_{\text{BL}}$, while $g_{st,a}$ represents the series of stomatal and atmospheric conductances; the stomatal conductance, $g_{st}$, is defined according to the multiplicative formulation by \cite{jarvis1976interpretation}, and in turn is affected by photosynthetically active radiation (PAR), the vapor pressure deficit (VPD), air temperature ($\theta_{\text{BL}}$), and leaf water potential ($\psi_l$), given in Eq.~\eqref{eq: jarv}. The atmospheric conductance depends on wind speed and surface roughness \citep{porporato2022ecohydrology,kaimal1994atmospheric} and is assumed to be constant here. The surface temperature ($\theta_l$) is obtained from the surface (leaf) energy balance of Eq.~\eqref{eq: surfeng}. 

\subsection{Hydraulic–Stomatal Coordination and Trade-Offs}

Soil-plant hydraulics and stomatal behavior regulate transpiration and, thus, the balance between latent and sensible heat fluxes. Plant hydraulic traits and physiological responses to water availability and stress vary among species \citep{anderegg2016meta}; typically, different strategies result in coordination among hydraulic and photosynthetic traits \citep{manzoni2013biological,matthews2024multiple}, resulting in species-dependent surface-flux partitioning patterns. More specifically, increased xylem vulnerability to drought sharply decreases $g_{p}$, the plant hydraulic conductivity, as soil water potential becomes more negative. In turn, stomata begin to close in response to dehydration, reducing $g_{st}$ at a rate that depends on species, thereby limiting transpiration into the atmosphere. The coordination of such traits, i.e., stem hydraulic conductivity and stomatal sensitivity, allows for a dynamic surface boundary with heat and moisture fluxes characteristic of diverse vegetated ecosystems. Overall, plant responses to water stress fall along a continuum between drought-avoidant (isohydric) and drought-resistant (anisohydric) strategies; these terms are often used to describe plant adaptation to drought, referring to tight or loose regulation of plant water potential, respectively \citep{klein2014variability, chen2019prediction}.

To account for such differences in plant adaptation in response to water availability, we consider a linear, positive relationship between stem water potential at 50$\%$ loss of stem hydraulic conductivity and the water potential at stomatal closure ($\psi_\mathrm{close}$) due to water stress,
\begin{equation}
\psi_{\mathrm{close}} = a\psi_{50,\mathrm{stem}} - b,
\label{eq: closeeq}
\end{equation}
where the parameters $a$ and $b$ reflect the functional form of the hydraulic safety margin across species (i.e., $|\psi_{50,\mathrm{stem}}| - |\psi_{\mathrm{close}}|$), which may vary depending on the plant functional group analyzed. Generally, we take $|\psi_{50,\mathrm{stem}}| > |\psi_{\mathrm{close}}|$ for isohydric and $|\psi_{50,\mathrm{stem}}| < |\psi_{\mathrm{close}}|$ for anisohydric species. In this work, the parameterization is taken from data on temperate broadleaf species \citep{chen2019prediction}. The difference between the water potentials driving these conductances defines a stomatal safety margin, which more fully describes both stomatal and xylem adjustments to drought conditions \citep{skelton2015predicting} and may change in coordination with other adaptive safety measures \citep{manzoni2013biological}. However, rather than focusing solely on distinct anisohydric and isohydric species, we further examine how trade-offs between stem conduction and stomatal sensitivity to declining water availability affect the overall water balance.

Accordingly, the connection between Eq.~\eqref{eq: closeeq} and the magnitude and timing of transpiration fluxes is provided by the relationship between stomatal conductance and leaf water potential.  A power-law relationship between maximum stomatal conductance and $\psi_{50,\text{close}}$ (leaf water potential at 50$\%$ of stomatal closure due to water stress) is presented based on changes in stomatal density and size \citep{henry2019stomatal}. The fitted relationship is given as: 
\begin{equation}
g_{st,\max } \propto\left|\psi_{50,\text{close} }\right|^{b_1}
\label{eq: gmaxeq}
\end{equation}
where $b_1$ is a fitted parameter from ten \textit{Ficus} species using standard major axes for log-transformed data as described by \cite{henry2019stomatal}. This relationship illustrates that as stomatal density increases and size decreases, leaves become more sensitive to closure but exhibit higher stomatal conductance when hydrated. The piecewise curve for $g_{st}$ vs. $ \psi_{l}$ is within the multiplicative function of \cite{jarvis1976interpretation}, described in Appendix \ref{app: spmodel}. For climate applications, this framework allows for natural scaling at the regional level. These physiological constraints may be interpreted as intrinsic averages of spatial distributions over a certain length scale representing the dominant functional plant type in a representative catchment or area of interest \citep{bartlett2025stochastic}. Altogether, factoring in physiological plant complexity into an otherwise atmospheric framework enables more complete land–atmosphere coupling. Here, we consider a range of $\psi_{50, \text{stem}}$ values to represent varying vulnerabilities and stomatal responses. 
\begin{table}[h!]
\centering
\caption{Key variables and symbols used in the model.}
\small
\setlength{\tabcolsep}{6pt}
\renewcommand{\arraystretch}{0.9}
\begin{tabular}{ll}
\hline
\textbf{Symbol} & \textbf{Definition} \\
\hline

$g_{st}$ & Stomatal conductance (m s$^{-1}$) \\
$g_{st,\max}$ & Maximum theoretical stomatal conductance (m s$^{-1}$) \\
$g_{p}$ & Plant (xylem) conductance (m s$^{-1}$ MPa$^{-1}$) \\
$g_{sr}$ & Soil–root conductance (m s$^{-1}$ MPa$^{-1}$) \\
$\psi_{s}$ & Soil water potential (MPa) \\
$\psi_{l}$ & Leaf water potential (MPa) \\
$\psi_{50,\text{close}}$ & Leaf water potential at which stomatal conductance declines to 50\% of $g_{st,\max}$ (MPa) \\
$\psi_{50,\text{stem}}$ & Water potential at 50\% loss of xylem hydraulic conductivity (MPa) \\

$g_{a}$ & Atmospheric conductance (m s$^{-1}$) \\
$q_{l}$ & Leaf specific humidity (kg kg$^{-1}$) \\
$q_{\text{sat}}$ & Saturation specific humidity (kg kg$^{-1}$) \\

$h$ & ABL height (m) \\
$\theta_{v,\text{BL}}$ & ABL virtual potential temperature (K) \\
$q_{\text{BL}}$ & ABL specific humidity (kg kg$^{-1}$) \\
$\gamma_{\theta_{v,f}}$ & Free-atmospheric temperature lapse rate (K m$^{-1}$) \\
$\gamma_{q,f}$ & Free-atmospheric humidity lapse rate (kg kg$^{-1}$ m$^{-1}$) \\
$\theta_{v,f,0}$ & Initial free-atmospheric potential temperature (K) \\
$q_{0,f}$ & Initial free-atmospheric specific humidity (kg kg$^{-1}$) \\

$z_{\text{LCL}}$ & Lifting condensation level (m) \\
$z_{\text{LFC}}$ & Level of free convection (m) \\
$z_{\text{LNB}}$ & Level of neutral buoyancy (m) \\
$\text{CAPE}$ & Convective available potential energy (J kg$^{-1}$) \\

$s$ & Relative soil moisture (-) \\
$Z_r$ & Rooting depth (mm) \\
$n$ & Soil porosity (-) \\
$\lambda$ & Rainfall arrival rate (day$^{-1}$) \\
$\alpha$ & Mean rainfall depth (mm) \\
$b$ & Soil retention parameter related to pore size (-) \\
$\overline{\psi}_s$ & Soil retention parameter (MPa) \\
$LAI$ & leaf area index (m$^{2} $m$^{-2}$) \\

\hline
\end{tabular}
\end{table}

\begin{figure}[h]
    \centering
    \begin{subfigure}{\textwidth}
        \centering
        \includegraphics[width=\linewidth]{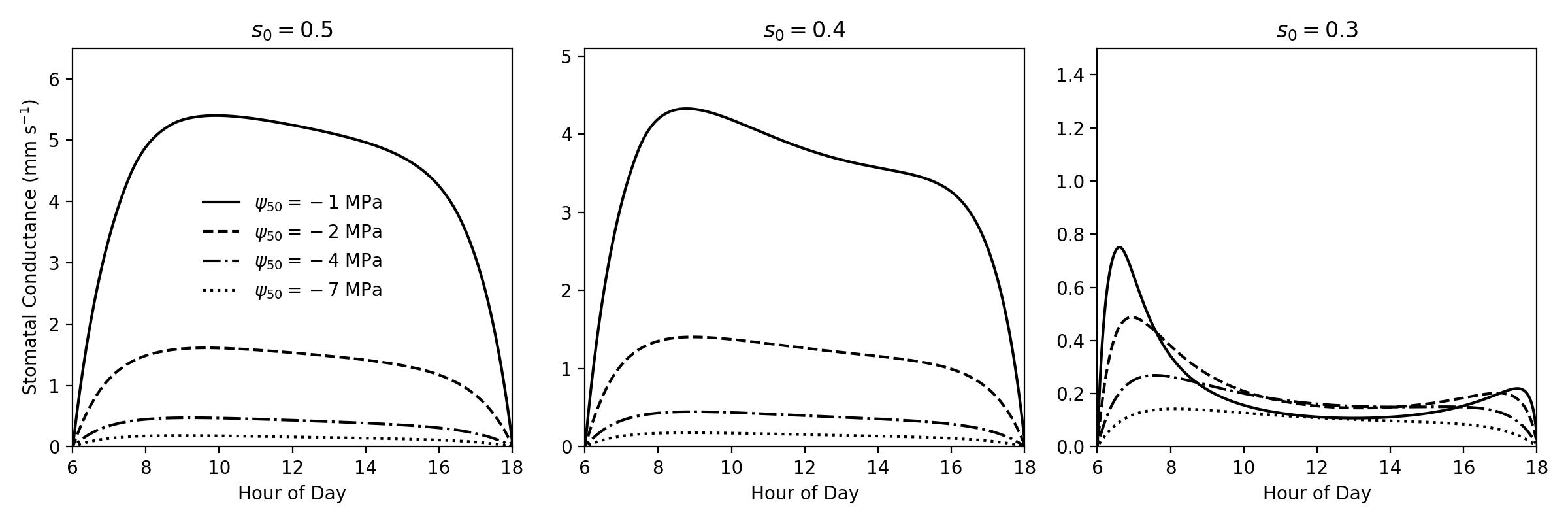}
        \label{fig:Stom1}
    \end{subfigure}
    \begin{subfigure}{\textwidth}
        \centering
        \includegraphics[width=\linewidth]{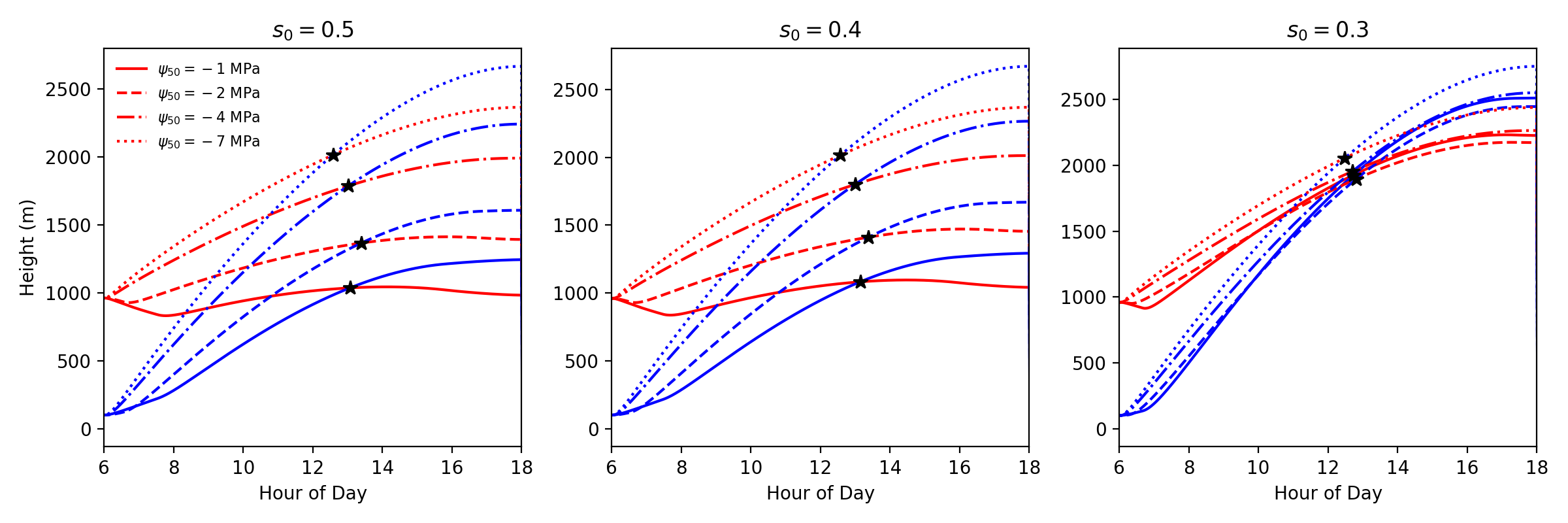}
        \label{fig:Stom2}
    \end{subfigure}
    \begin{subfigure}{\textwidth}
        \centering
        \includegraphics[width=\linewidth]{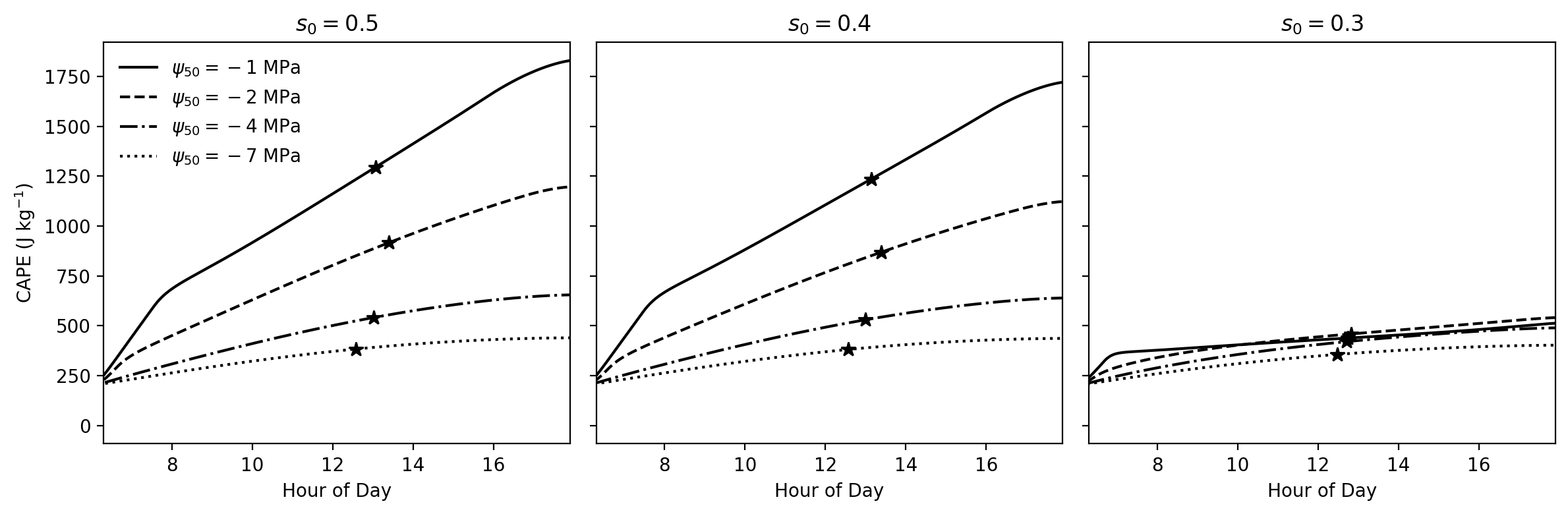}
        \label{fig:Stom3}
    \end{subfigure}
    \caption{Diurnal evolution of stomatal conductance, LCL height (red lines), ABL height (blue lines), and CAPE throughout the day for three initial relative soil moisture conditions: 0.5, 0.4, and 0.3, from left to right. The atmospheric parameters are taken from \cite{yin2015land} based on early morning measurements at the Central Facility, Southern Great Plains. The plant parameters (i.e., $LAI$, $Z_r$, and $g_a$) are taken from the Forest tab in Table \ref{tab:plant_traits}, given a sandy loam soil. The black stars represent the crossing time of the LCL and ABL.}
    \label{fig:sidebyside}
\end{figure}

\section{Diurnal Evolution}
 
 We first analyzed the system at a subdaily scale by tracking the temporal evolution of stomatal conductance, ABL height, LCL height, and CAPE from sunrise to sunset. The diurnal dynamics of each state variable were resolved using radiative forcing, modeled as a parabolic function with zero input at night. Starting the simulation with three different soil moisture conditions at midnight (Figure \ref{fig:sidebyside}), we ran the full land-atmosphere system of equations with different $\psi_{50,\text{stem}}$ values. We found that the fluxes of water vapor show varying diurnal progression at all soil moisture levels for different plant water use strategies, given different initial conditions for the relative soil moisture ($s_0$), ranging from wet to dry. The LCL height was obtained using the analytical expression in Appendix \ref{app: LCLh}. The crossing time between the LCL (red lines) and ABL height (blue lines) is marked with a black star. When water availability is high ($s_0 = 0.5$), plants with high maximum stomatal conductance humidify the local boundary layer through a high transpiration rate, decreasing the height of the LCL and ABL, producing high CAPE. In the intermediate soil moisture range, plants with sustained stomatal opening but a low maximum stomatal conductance (less negative $\psi_{50,\text{stem}}$) still may produce high enough CAPE to trigger convection, although transpiration is, on average, lower during the day. However, plants with high maximum stomatal conductance still maintain high CAPE values in this intermediate range, which may contribute to more severe precipitation. As the soil dries and the environment leaves the ``transition zone" during the evolution of a dry-down (from $s_0 = 0.4$ to $s_0 = 0.3$), the differences in the LCL-ABL start to converge, and transpiration is maintained at a higher level during the daylight hours for plants with lower maximum stomatal conductance. However, we see that CAPE across all plant water-use strategy scenarios converges when soil water is limited. 

These findings, however, do not fully capture climate feedbacks, as transpiration differences between plant types are relatively small at the same soil moisture level. Overall, there exists two timescales that govern the triggering of moist convection: (1) local, sub-daily turbulent and moisture fluxes from the surface drive ABL and LCL growth, altering their crossing time and CAPE at initiation, described in this section. Not yet considered in Figure \ref{fig:sidebyside} is (2), the daily evolution of soil moisture through drying (i.e., negative drift), which is modified by surface characteristics and subsequently alters precipitation statistics, creating a feedback loop.

\section{Dry-down}

Given the above contributions, we then performed a full dry-down simulation ($I_T = 0$) for a range of plant water-use strategies and soil types. At nighttime, transpiration is assumed to be negligible due to stomatal closure. We assumed a collapsed, shallow boundary layer at night (denoted $h_0$) or after rainfall until nighttime. For the linear profiles of the free-atmospheric lapse rates, we used parameters from \cite{yin2015land} as a representative baseline to examine the effects of the land-surface boundary on convective triggering. The parameters are derived from early-morning observations at the Central Facility, Southern Great Plains (CF-SGP), on 18 July 2009.
We initialize the model with $q_{\text{BL}}$ = $q_{0,f}$ ($\text{kg} \ \text{kg}^{-1}$) and $\theta_{v,\text{BL}}$ = $\theta_{v,f,0}$ ($\text{K}$).

\begin{figure}[h]
\centering

\includegraphics[width=\textwidth]{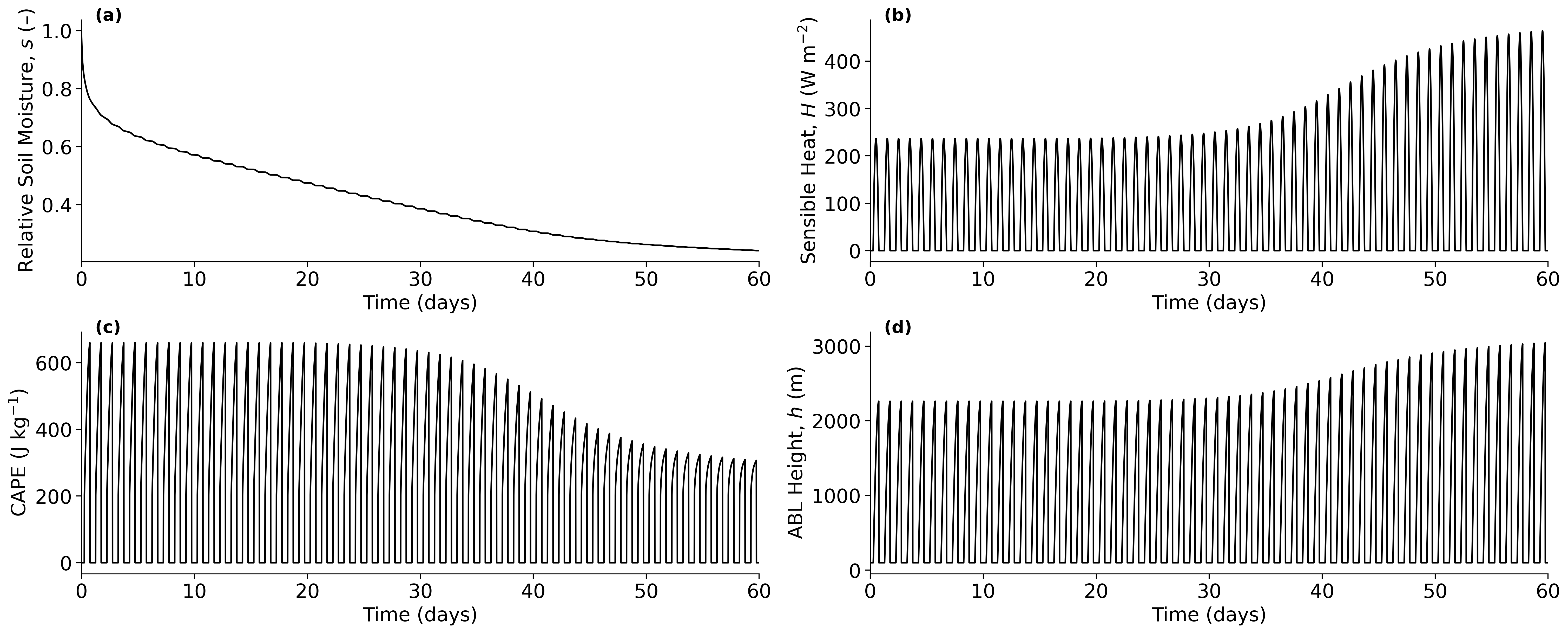}

\caption{Example relative soil moisture ($s$), sensible heat ($H$), CAPE, and boundary layer height evolution for a full dry-down beginning at $s=1$. The parameter values are the same as in Figure \ref{fig:sidebyside}, with a $\psi_{50,\text{stem}}$ = -4 MPa.}. 
\label{fig:realdrydown}
\end{figure}

\begin{figure}[h]
    \centering
    
    \begin{subfigure}{0.49\textwidth}
        \centering
        \includegraphics[width=\linewidth]{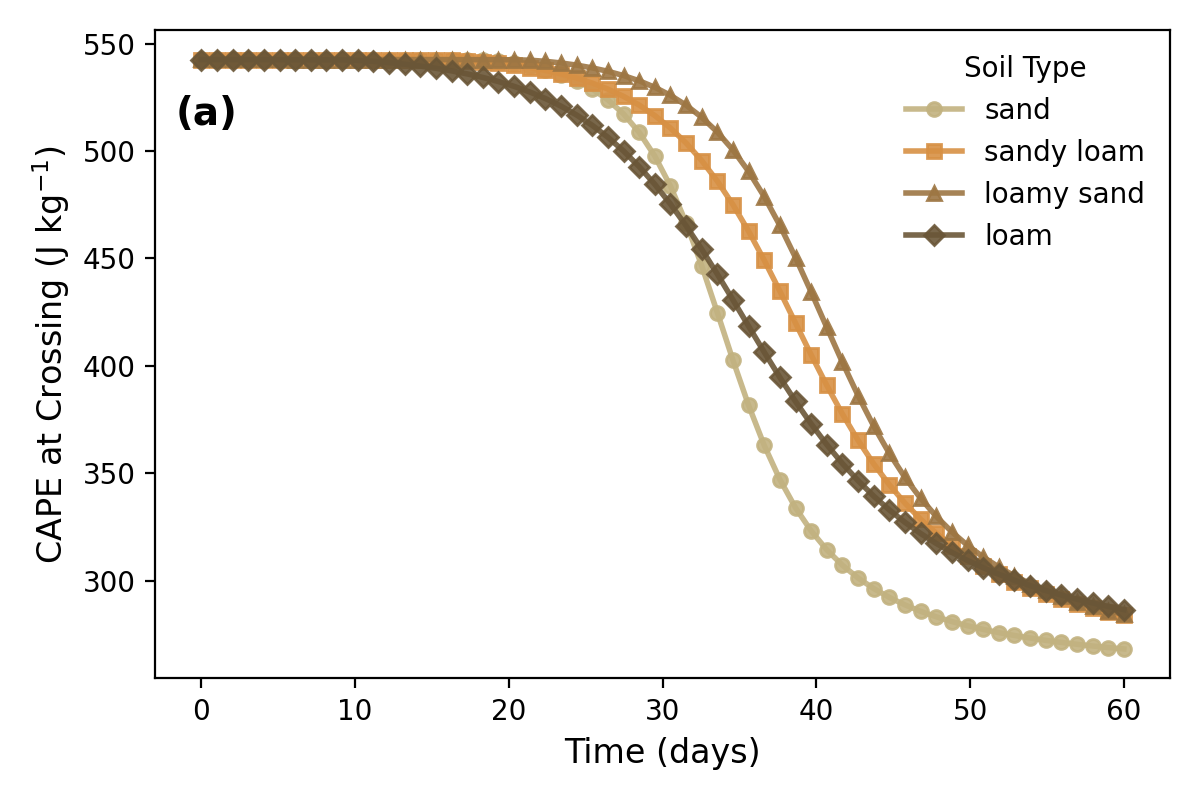}
    \end{subfigure}
    \hfill
    \begin{subfigure}{0.49\textwidth}
        \centering
        \includegraphics[width=\linewidth]{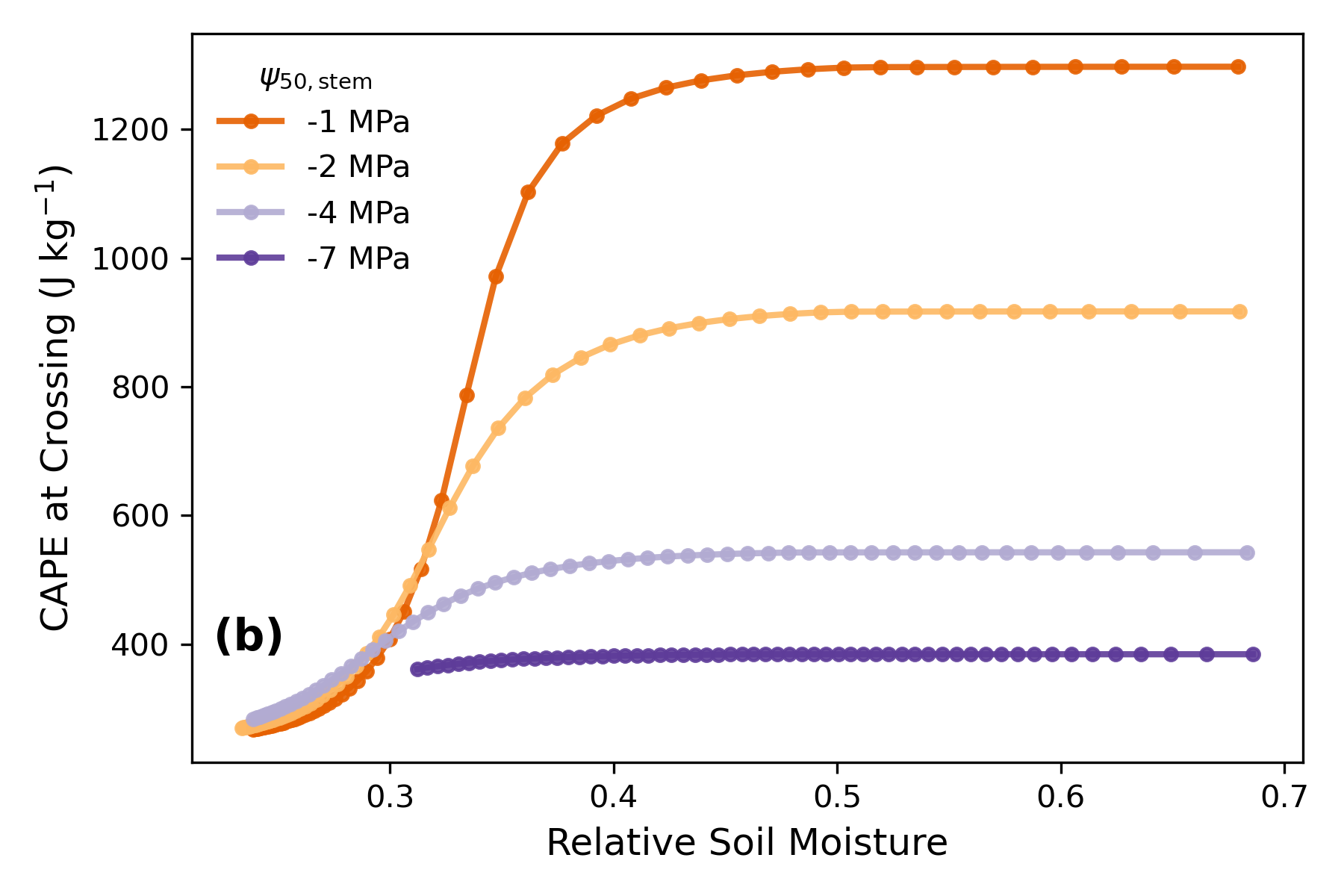}
    \end{subfigure}
    
    \vspace{0.5em}
    
    \begin{subfigure}{0.49\textwidth}
        \centering
        \includegraphics[width=\linewidth]{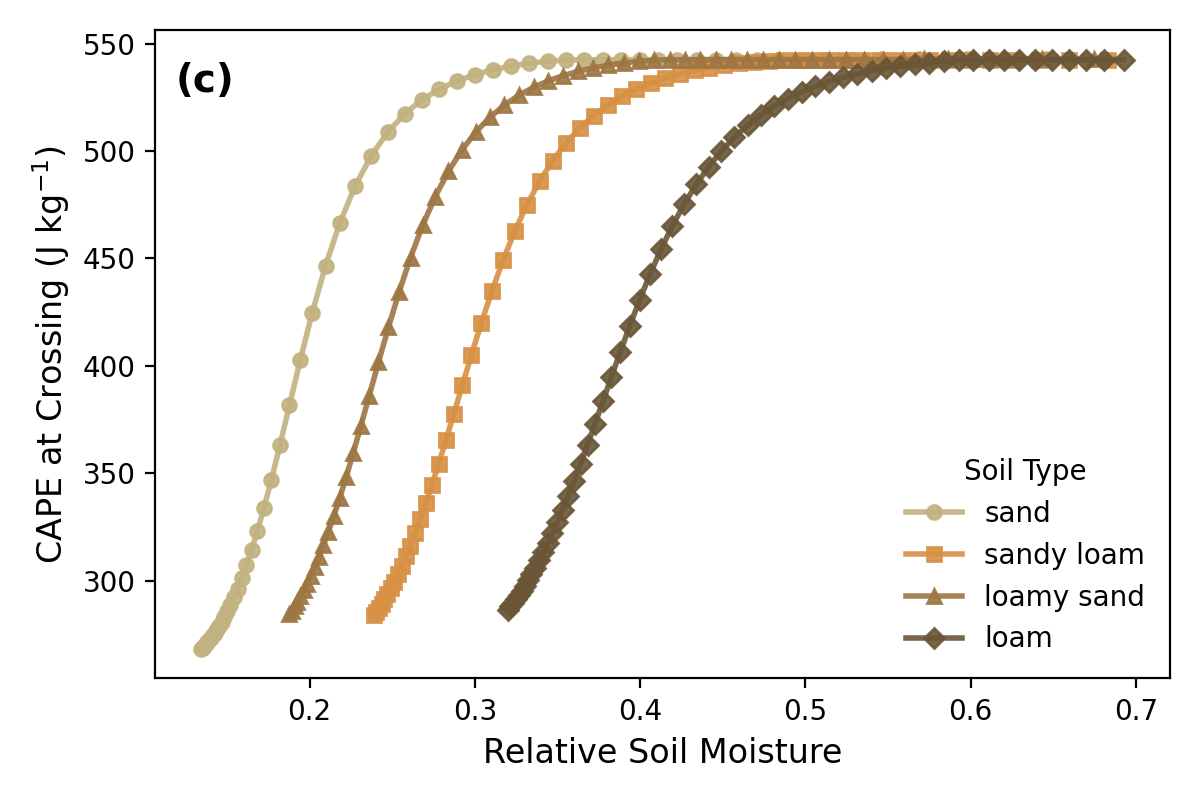}
    \end{subfigure}
    \hfill
    \begin{subfigure}{0.49\textwidth}
        \centering
        \includegraphics[width=\linewidth]{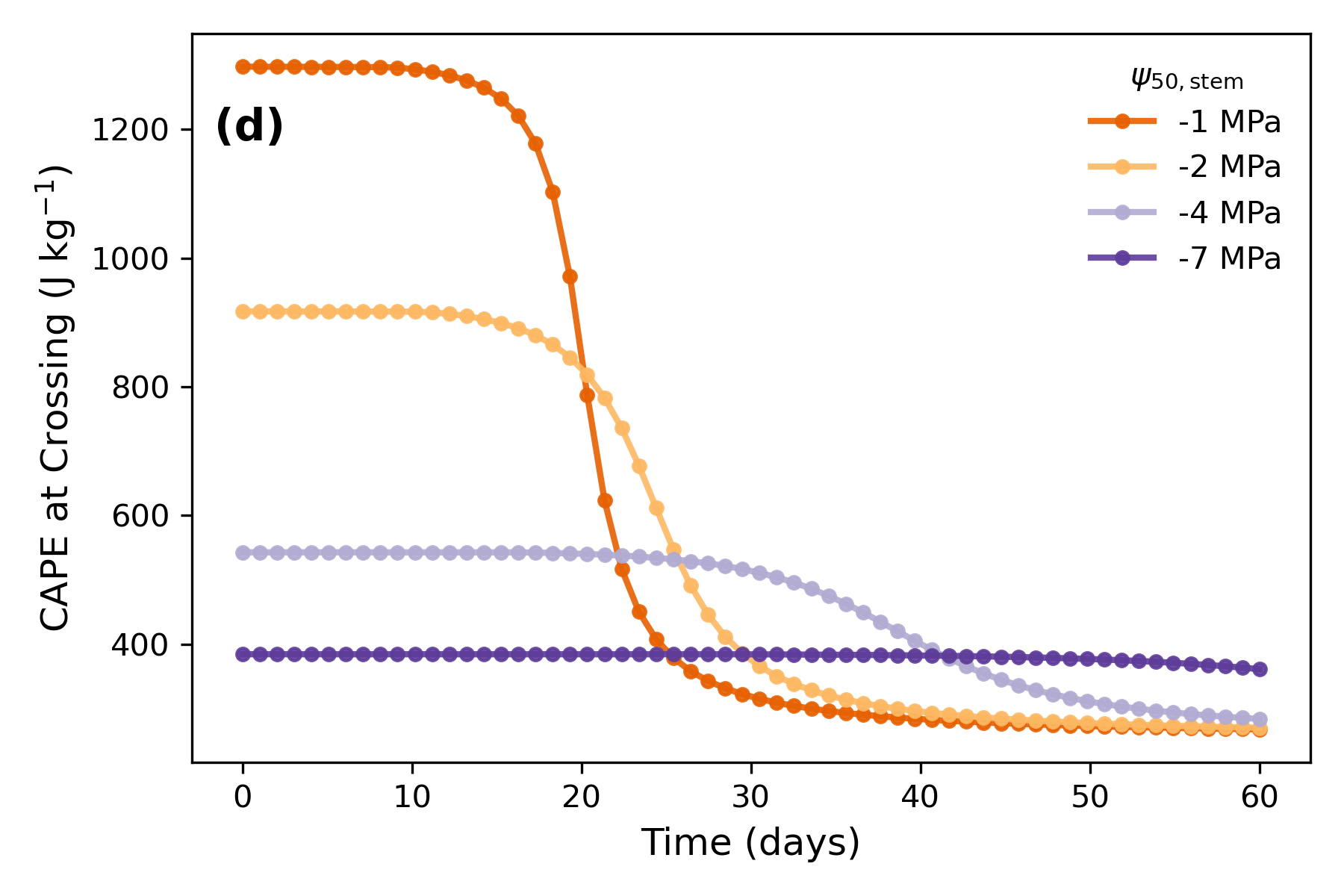}
    \end{subfigure}

    \caption{Panel (a) represents CAPE at crossing time vs.\ day of dry-down for different soil types. Panel (b) illustrates the same dry-down for four different $\psi_{50,\text{stem}}$ values, given the same parameters as Figure \ref{fig:realdrydown}, other than the modification of $\psi_{50,\text{stem}}$ and soil type. Panels (c) and (d) represent CAPE at crossing vs. soil moisture, ($s$).}
    \label{fig:soiltype}
\end{figure}

Figure \ref{fig:realdrydown} illustrates the evolution of key variables during a 60-day dry-down, showing the progressive decline in soil moisture, the corresponding rise in sensible heat flux, and the reduction in CAPE as the atmospheric boundary layer (ABL) deepens. Although soil moisture and CAPE are not one-to-one on the sub-daily scale, soil moisture still exhibits a functional relationship with the magnitude of CAPE from one day to the next (i.e., a daily scale). We capture the described functional relationship by extracting both the LCL-ABL crossing time (hour relative to sunrise) and the CAPE at crossing time, as discussed in the previous section. By doing this, we extend the work of \cite{daly2004coupled} by forming an upscaled, daily relationship between soil moisture and CAPE, among other atmospheric variables, that can be parameterized by soil and plant functional type. These CAPE at LCL crossing curves for plant and soil type are shown in Figure \ref{fig:soiltype}.

Given such conditions, we compare the evolution of key hydrologic and atmospheric variables in response to drought across a range of soil types and, separately, across multiple $\psi_{50,\text{stem}}$ values (i.e., different plant water-use strategies). We find that, for the relative soil moisture values considered, the CAPE at the crossing time generally increases from loam to sand, but that soil moisture decreases faster over a dry-down for sand, as in panel (a) of Figure \ref{fig:soiltype}. Changing plant type, or modifying different values of $\psi_{50,\text{stem}}$, we find that CAPE at crossing maintains a higher value over the dry-down period, but a lower maximum when soil moisture is high, as the $\psi_{50,\text{stem}}$ becomes more negative, as shown in panel (b) of Figure \ref{fig:soiltype}. Plants with low maximum conductance and thus a highly negative $\psi_{50,\text{stem}}$ values may contribute to more frequent storms (i.e. positive feedback), as CAPE is held high enough for convection long after the soil is wet from a storm event, as moisture is removed more slowly from the soil, and therefore the plant. Still, the maximum plant conductance must be high enough for CAPE to reach the minimum threshold. When soil moisture falls below field capacity (i.e., a water-limited regime), transpiration dynamics are constrained by soil moisture, tightening the coupling between plant water use and soil moisture. The ABL is, in part, regulated by the strength and dynamics of the transpiration flux. As a result, the surface hydraulic state (as reflected in changes in water potential in both the plant and soil) plays a major role in convective initiation. Plant parameters also influence convection. For example, modifying the leaf area index ($LAI$) changes the sigmoidal curves in Figure \ref{fig:soiltype}. Because $LAI$ regulates the magnitude of water vapor flux, reducing $LAI$ significantly lowers CAPE at crossing time for moderate to high soil moisture levels. Consequently, plants with low maximum conductance and low $LAI$ may never generate sufficient CAPE for convection. Given this, and considering the variation in plant responses, we observe that multiple combinations of factors govern the timing and magnitude of convective triggering. However, the relationship between soil moisture and crossing time depends strongly on atmospheric stability and humidity, which may be explored in the future through temporally dynamic free-atmospheric lapse rates. Here we consider static lapse rates to isolate the role of surface fluxes in convective triggering; however, in the sections that follow, we will examine the free-atmosphere latitudinal dependence on convection.  

Importantly, altering the 50$\%$ loss of conductivity also changes the shape of the CAPE–soil moisture relationship. Although the overall sigmoid shape is preserved, the slope and corresponding CAPE magnitude decrease for plants that continue water use at lower soil moisture levels. In spite of that, increased soil water reserves during the latter half of the dry-down period provide sustained fuel for convection in plants that continue to transpire during the dry periods, thereby enhancing the likelihood of rainfall through convective triggering. Collectively, these two simulations highlight the importance of dynamic plant and soil water availability in determining the system's thermodynamic state, rather than static moisture alone. 

\section{Rainfall Feedback and Long-Term Soil Moisture Dynamics}
To investigate the long-term impact of the connection between CAPE and the LCL crossing on soil moisture and land-surface processes, we link these variables to the likelihood of convection. The resulting feedback loop drives a stochastic rainfall model, which, when integrated with the land-atmosphere system, enables exploration of the main pathways of interaction between the changing land surface and the buildup of CAPE and precipitation on seasonal and multi-year timescales.  

\subsection{Stochastic Rainfall Model with CAPE and Soil-Moisture Feedback}

As is typical in stochastic ecohydrology \citep{porporato2022ecohydrology}, we model precipitation as instantaneous storm events. We consider pulses from both convective and stratiform systems (considered as background), the former of which is dependent on the soil moisture state through CAPE, i.e.,
\begin{equation}
P(t) = \sum\limits_{i=1}^{N(t)} P_c\left(\text{CAPE}^*\right)\delta(t-t_i)+\sum\limits_{j=1}^{M(t)} P_0\delta(t-t_j).
%P(t) = P_c(\text{CAPE}[s(t^*)])+P_{0}(t).
\label{totalprecip}
\end{equation}
Here, the star indicates the variables at the time, $t^{*}$, of the LCL-ABL crossing, and the CAPE-dependent convective precipitation amounts, $P_c\left(\text{CAPE}^*\right)$, and the stratiform precipitation amount, $P_0$, are instantaneous pulses of random depth extracted from exponential distributions at the respective times $\{t_i\}(i=1,2,3,...)$ and $\{t_j\}(j=1,2,3,...)$, as indicated by Dirac delta function, $\delta(\cdot)$, with $N(t)$ and $M(t)$ counting random rainfall occurrences. This defines the sequence $t_i$. Both the convective rainfall and stratiform rainfall processes, $P_c$ and $P_0$, are reasonably represented by marked Poisson processes, which are defined by exponentially distributed interarrival times, $t_j-t_{j-1}$. The convective precipitation is represented as a non-homogeneous Poisson process with frequency $\lambda_c(s)$, which occurs when the ABL crosses LCL and CAPE is above 400 J kg$^{-1}$ \citep{yin2015land}, while the stratiform precipitation is represented as a homogeneous Poisson process with a constant rate. For each stratiform event at rate $\lambda_0$, the total rainfall `mark', $P_0$, is drawn from an exponential distribution of mean depth $\alpha_{0}$, and for the convective events at rate $\lambda_c(s)$, the rainfall total `mark', $P_c$ is also drawn from an exponential distribution.

The total convective precipitation depth consists of a component derived from the moisture budget aboves the boundary layer complemented by a lower-boundary-layer moisture contribution, $M_\text{BL}$,
\begin{equation}
\label{eq: combo}
P_c = \hat{P}_c + \varepsilon_{\mathrm{BL}}\, q_{\mathrm{BL}}\, h^{*},
\end{equation}
where $\hat{P}_c$ is the precipitation depth generated above the lifting condensation level and is treated as a random variable drawn from the distribution $p_{\hat{P}_c}(\hat{P}_c)$. The second term represents the fraction of vapor stored in the atmospheric boundary layer of depth $h^*$ that precipitates during the event, with $\varepsilon_{\text{BL}}$ being the fraction of ABL water vapor contributing to precipitation. On the daily time scale, this rainfall is assumed to occur instantaneously.

The convective rainfall depth from the moisture budget above the ABL is the product of the rainfall intensity, $I_c$, storm duration, $T_c$,
\begin{equation}
\hat{P}_c = I_c T_c,
\label{eq:Tc}
\end{equation}
where $I_c$ is derived from the atmospheric moisture budget under the conditions present at the time of convection, and $T_c$ is exponentially distributed with mean $\delta$ [T] \citep{eagleson1978climate, may2007statistical, sangiorgio2020spatio}:
\begin{equation}
p_{T_c}(T_c)=\frac{1}{\delta}e^{-T_c/\delta}.
\label{eq:p_hc}
\end{equation}
This formulation explicitly links the moisture budget to convective rainfall while capturing the stochastic variability of storm duration, although alternative distributions could be adopted depending on regional storm characteristics. Performing a change of variables with Eqs. (\ref{eq:Tc}) and (\ref{eq:p_hc}), the probability density of rainfall depth, $\hat{P}_c$ is
\begin{equation}
p_{\hat{P}_c}(\hat{P}_c)=\frac{1}{I_c} p_{T_C} (\hat{P}_c/I_c),
\end{equation}
which implies that $\hat{P}_c$ also follows an exponential distribution with mean $\mathbb{E}[\hat{P}_c] = I_c{\delta}$.

The storm intensity, $I_c$, follows from a pseudo-steady-state atmospheric moisture budget evaluated at the time of convection $t^*$ \citep{muller2011intensification}.  This yields (see Appendix \ref{app:PI APP} for details),

\begin{equation}
\label{eq: above abl}
I_c \equiv
\int_{z^{*}_{\mathrm{LCL}}}^{z^{*}_{\mathrm{LNB}}}
-\varepsilon \frac{dq_{\mathrm{sat}}}{dz}\, w^{*}(z)\, \rho_a(z)\, dz,
\end{equation}
where the integral represents condensation associated with the vertical transport of saturated moisture between the lifting condensation level, $z^{*}_{\mathrm{LCL}}$, and the level of neutral buoyancy, $z^{*}_{\mathrm{LNB}}$, which is based on the change in saturated specific humidity with height, $\frac{dq_{\text{sat}}}{dz}$, the updraft speed $w^*(z)$, and air density $\rho_a(z)$. The precipitation efficiency factor, $\varepsilon$ is defined in Eqs.~\eqref{eq: effic} and~\eqref{eq: efficIC}. The maximum updraft speed at any height $z$ \citep{williams2017meteorological} in Eq.~\eqref{eq: above abl} can be expressed as:
\begin{equation}
\label{eq: updraft}
w^{*}(z) = \sqrt{2 \ \text{CAPE}(z)}.
\end{equation}
The precipitation efficiency factor $\varepsilon = \varepsilon'\sqrt{\eta}$ in Eq.~\eqref{eq: above abl}, however, includes an efficiency factor $\eta$, accounting for the dilution due to entrainment (mixing) of environmental air that reduces overall buoyancy \citep{peters2020formula} along with cooling involved in liquid-water loading.

\begin{figure}[h]
\centering
\includegraphics[width=\textwidth]{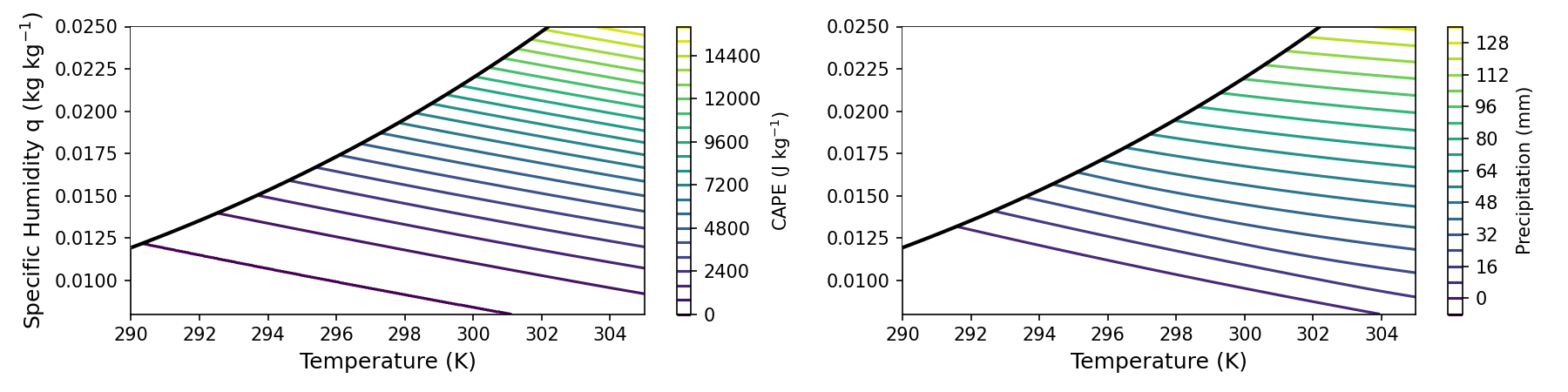}
\caption{CAPE and convective event precipitation as a function of surface temperature and specific humidity values shown by the contour plot, given that the ABL has just reached the LCL, given an average storm duration time of 20 minutes, and precipitation efficiency, $\varepsilon$ = $\varepsilon_{\text{BL}}$, of 0.1.}
\label{fig: CAPEPRECIP}
\end{figure}

\subsection{Long-Term Dependence of Rainfall Frequency and Intensity on Soil Moisture} 

The long-term effects of land-surface regulation on the triggering of convective rainfall are made clear by the state dependence of both precipitation frequency and intensity. To explore this dependence, we ran long-term simulations of the coupled soil moisture/CAPE/rainfall dynamics and reconstructed multi-day averages of rainfall frequency and corresponding relative soil moisture. In this simulation, we averaged the rainfall frequency over a 50-day period, denoted as $\overline{\lambda} \  ( \text{day}^{-1})$, while taking the corresponding soil moisture on the first day of this period.

\begin{table}[h]
\centering
\caption{Plant hydraulic traits and seasonally averaged free atmospheric parameters at two latitudes.}
\resizebox{\textwidth}{!}{%
\begin{tabular}{lcccc|cccc|cccc}
\toprule
 & \multicolumn{4}{c}{Plant Traits} & \multicolumn{4}{c}{37$^\circ$N} & \multicolumn{4}{c}{43$^\circ$N} \\
\cmidrule(lr){2-5} \cmidrule(lr){6-9} \cmidrule(lr){10-13}
Cover &
$\psi_{50,\text{stem}}$ &
$Z_r$ &
$g_a$ &
$LAI$ &
$\gamma_{q,f}$ &
$\gamma_{\theta_{v,f}}$ &
$\theta_{v,f,0}$ &
$q_{0,f}$ &
$\gamma_{q,f}$ &
$\gamma_{\theta_{v,f}}$ &
$\theta_{v,f,0}$ &
$q_{0,f}$ \\
 &
[MPa] &
[mm] &
[m s$^{-1}$] &
[m$^{2}$ m$^{-2}$] &
[kg kg$^{-1}$ m$^{-1}$] &
[K m$^{-1}$] &
[K] &
[kg kg$^{-1}$] &
[kg kg$^{-1}$ m$^{-1}$] &
[K m$^{-1}$] &
[K] &
[kg kg$^{-1}$] \\
\midrule
Grass  & -2 & 300 & 0.01 & 2
& $-2.6 \times 10^{-6}$ & 0.00414 & 300 & 0.0127
& $-2.4 \times 10^{-6}$ & 0.00441 & 296 & 0.0107 \\
Forest & -4 & 800 & 0.02 & 8
& $-2.9 \times 10^{-6}$ & 0.00454 & 299 & 0.0138
& $-2.4 \times 10^{-6}$ & 0.00499 & 295 & 0.0111 \\
\bottomrule
\end{tabular}%
}
\label{tab:plant_traits}
\end{table}

The rainfall–soil moisture dependence, and the transitions from wet to dry periods (or vice versa), change when land-surface properties vary. Given this coupling relationship, we considered the effects of different plant types (see Table \ref{tab:plant_traits}): short grass (cropland), with low atmospheric conductance, shallow rooting depth, and low $LAI$; and large trees, with higher atmospheric conductance \citep{kelliher1993evaporation}, characterized by more negative $\psi_{50,\mathrm{stem}}$ \citep{lens2016herbaceous}, deeper active rooting depth \citep{jackson1996global, schenk2002global} and high $LAI$. The free atmospheric parameters in the table, used in Eqs.~\eqref{eq: free lapse rate} and~\eqref{eq: free lapse rate humid}, are seasonally averaged values from \cite{cerasoli2021cloud}, taken from 2001 through 2010 at 37$^{\circ}$ North and 43$^{\circ}$ North, originally from European Centre for Medium-Range Weather Forecasts (ECMWF) coupled climate reanalyses of the 20th century (CERA-20C). The varying atmospheric parameters capture the location dependence of surface–atmosphere coupling through differences in free-atmosphere stability and moisture. As latitude increases, there is a marked decrease in the intercepts of the linear free atmospheric equations, namely $\theta_{v,f,0}$ and $q_{0,f}$, with an upward trend in $\gamma_{q,f}$ but no considerable trend in $\gamma_{\theta_{v,f}}$ \citep{cerasoli2021cloud}. The two latitudes are chosen to compare convective regimes in the southern and northern United States and to represent mid-latitude environments globally. Although synoptic forcing and other complex atmospheric processes influence the development of convective rainfall, we use free-atmospheric profiles to represent the mean growing-season climatological state. The simulation was run for a total of 10,000 days to achieve steady-state conditions. Subsequently, probability density functions (PDFs) of relative soil moisture and CAPE at crossing time were reconstructed. Finally, a scatter plot of average rainfall frequency, conditioned on soil moisture, was generated for each simulation.

To isolate the effects of vegetation cover on land–surface coupling, we first conducted simulations with identical free-atmosphere parameters for both cropland and forest, taken from \cite{yin2015land} based on measurements at the Central Facility in the Southern Great Plains. The corresponding time series reveal that changes in plant cover produce large differences not only in ensemble-average soil moisture, but also in climate persistence.

\begin{figure}[h!]
    \centering
    \begin{subfigure}{1\textwidth}
        \centering
        \includegraphics[width=\linewidth]{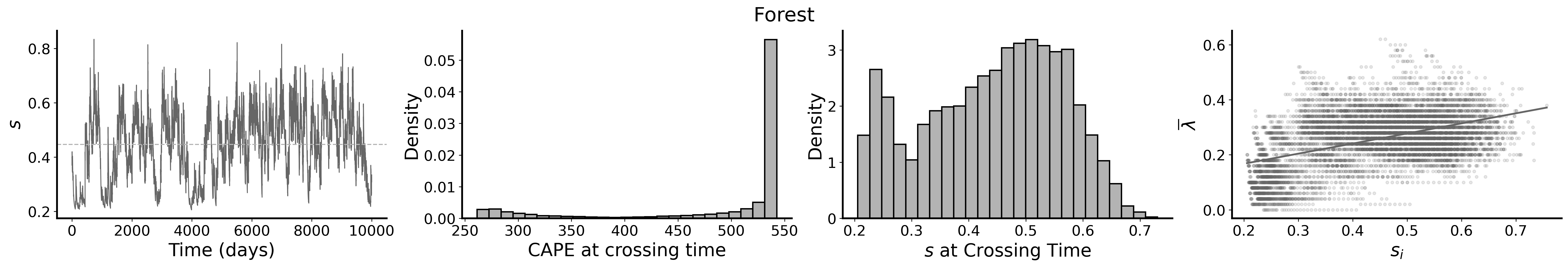}
        \label{fig:panelA}
    \end{subfigure}
    \hfill
    \begin{subfigure}{1\textwidth}
        \centering
        \includegraphics[width=\linewidth]{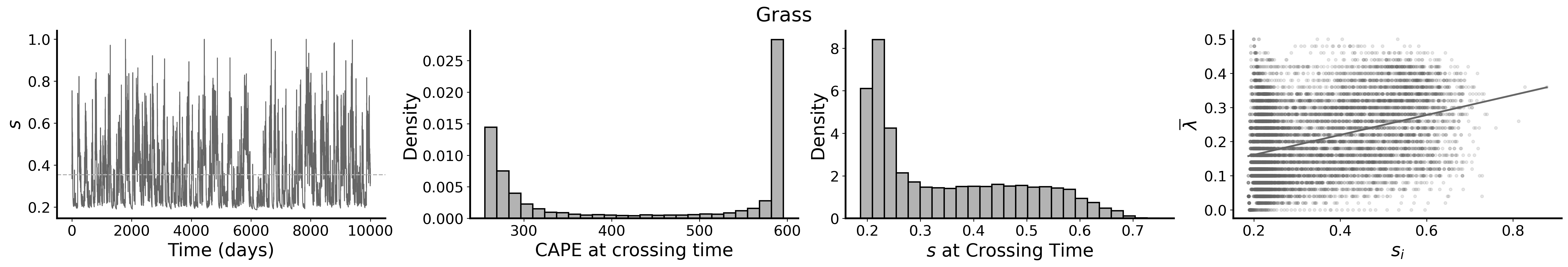}
        \label{fig:panelB}
    \end{subfigure}

    \caption{Grouped plot for each land cover type, showing time series of soil moisture, corresponding histograms, and scatter plot of averaged frequency over 50 days ($\bar{\lambda}$) versus initial soil moisture ($s_i$).}
    \label{fig: rainfallfreqgraph}
    
\end{figure}

As shown in Figure \ref{fig: rainfallfreqgraph}, there is a positive feedback relationship between the average rainfall frequency ($\bar{\lambda}$), conditioned on the soil moisture value ($s_i$), taken from the first day of a 50-day rolling window. Figure \ref{fig: rainfallfreqgraph} illustrates how wet soils sustain wet climates, while dry soils inhibit frequent convective rainfall. However, the timing and duration of such persistent events strongly depend on the dominant plant cover, as seen in the differences between the top and bottom rows of Figure \ref{fig: rainfallfreqgraph}. When comparing the two ground cover types, wet persistence is much more common for tree cover than for grass, although both exhibit long-term bimodal probability distributions of relative soil moisture and CAPE. In both simulations, the clustering of frequency values around wet and dry states arises from the formulation of the model, with dry periods clustering around the frontal precipitation frequency and wet periods dominated by the combined frequency of both $P_0$ and $P_c$ (when soil moisture is favorable for convection), set as $0.1$ and $0.2$ $\text{day}^{-1}$ in these simulations, respectively.

Previous studies have shown that soil and vegetation types directly affect the buildup of energy and moisture, altering the timing or occurrence of convective triggering \citep{eltahir1996precipitation, miralles2025vegetation}. This then relates directly to the results in Figure \ref{fig: rainfallfreqgraph}, where preference of soil moisture states exists over extended periods of time (i.e., up to multiple years), switching from dry to wet regions. This occurs most often in deep-rooted tree cover, but rarely in shallow-rooted grassland, where switching occurs frequently but is shorter-lived. The reason is directly due to plant water availability and water use strategies. The tight and rapid exchanges between the ABL and the surface, along with the coupling of rainfall intensity and near-surface moisture, produce rapid switching between wet and dry soil moisture states, which can persist for relatively long times. The switching frequency, however, depends on how moisture fluxes respond to soil moisture. On one hand, increased stomatal maximum conductance during wet periods increases the likelihood of intense and frequent rainfall, but the corresponding steep decline in transpiration over fewer days and the quick use of surface water stores will not promote the persistence of a wet state, as in Figure \ref{fig: rainfallfreqgraph}. Consequently, grasslands exhibit lower average soil moisture due to more rapid drying. We find that for species that maintain a steadier conductance at more negative stem water potentials, i.e., higher latent heat flux, wet persistence tends to last longer. The difference in vegetation cover may have large effects over multiple years, especially when climate change alters the biotic and abiotic landscape.

\begin{figure}[hbt!]
    \label{fig: rainfallfreqgraphlat}
    \centering

    \begin{subfigure}{1\textwidth}
        \centering
        \includegraphics[width=\linewidth]{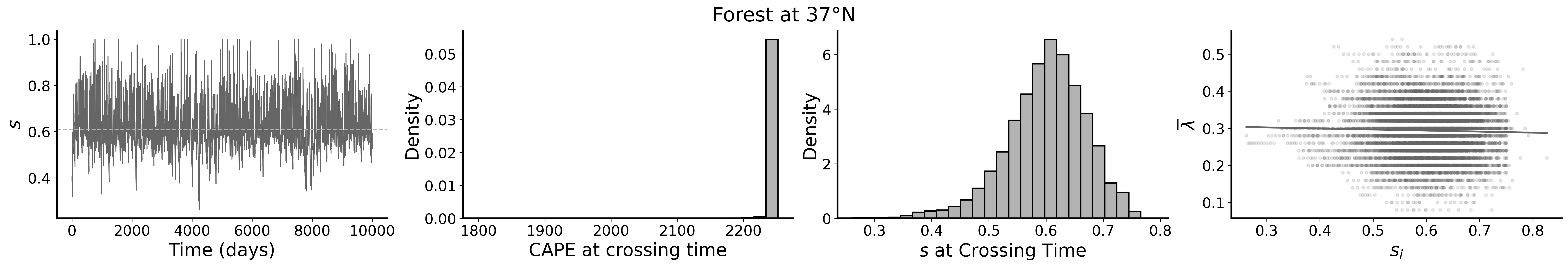}
        \label{fig:panelA}
    \end{subfigure}
    \hfill
    \begin{subfigure}{1\textwidth}
        \centering
        \includegraphics[width=\linewidth]{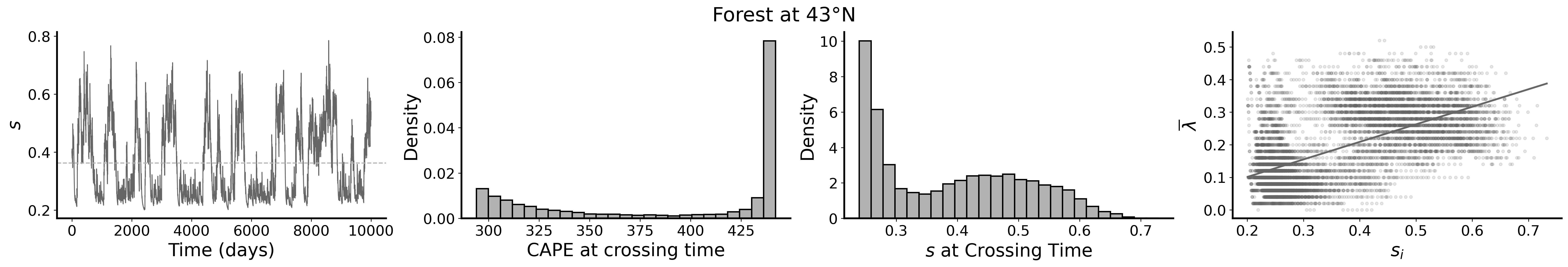}
        \label{fig:panelB}
    \end{subfigure}
        \begin{subfigure}{1\textwidth}
        \centering
        \includegraphics[width=\linewidth]{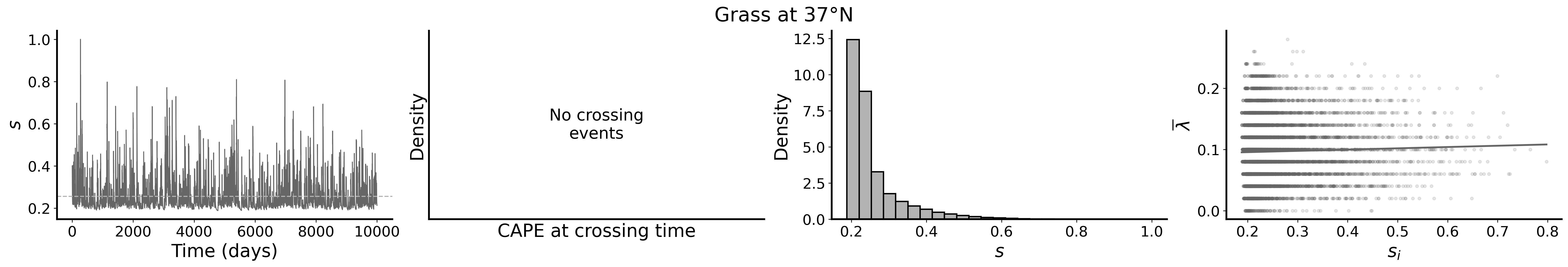}
        \label{fig:panelC}
    \end{subfigure}
        \begin{subfigure}{1\textwidth}
        \centering
        \includegraphics[width=\linewidth]{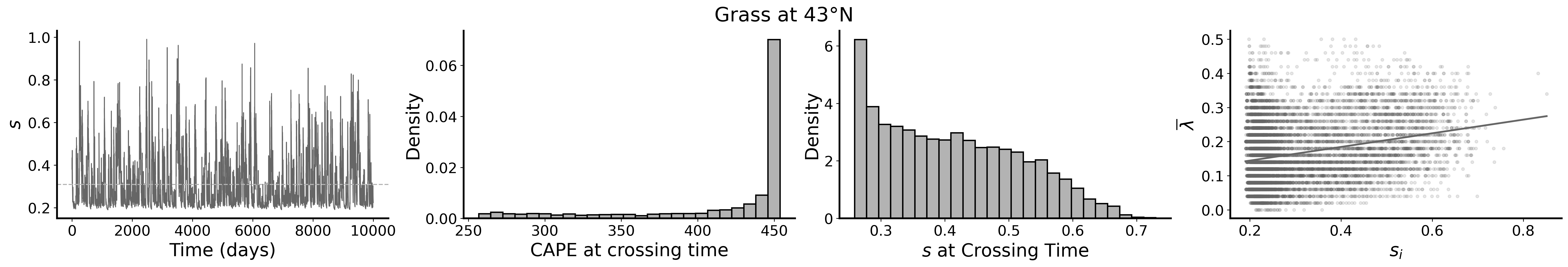}
        \label{fig:panelD}
    \end{subfigure}

    \caption{Latitude-dependent grouped plot for each landcover type, showing time series of soil moisture, corresponding histograms, and scatter plot of averaged frequency over 50 days ($\bar{\lambda}$) versus initial soil moisture ($s_i$). Importantly, grass/cropland at 37$^\circ$ has atmospheric conditions that do not produce crossing times; the soil moisture at crossing histogram, in this case, is the histogram of $s$.}
    \label{fig: rainfallfreqgraphlatitude}
\end{figure}

\subsection{Longitudinal Effects and Plant Functional Types}

We also investigated latitudinal changes in these feedbacks \cite{cerasoli2021cloud}, given also their importance in assessing the climate mitigation potential of reforestation and afforestation. Focusing on differences in plant functional types, we found that at higher latitudes, feedbacks from tree cover are significantly stronger than those from grass, with grass failing to generate sufficient CAPE to sustain convection or climate persistence. Conversely, forest land cover sustains high enough CAPE to favor convection and, thus, overall wetter conditions. The same mechanisms described above, including rapid soil drying from water-use strategies and low $LAI$/atmospheric conductance, decrease overall CAPE. Furthermore, the grassland temperature profile is slightly less unstable, and the soil dries more quickly, limiting convection in this framework. For forests, climate behavior at lower latitudes strongly favors convection and persistent wetness due to greater atmospheric instability and enhanced availability of lower-level moisture (as shown by CAPE–soil moisture curves in the dry-down section). However, even though grass at more southern latitudes favors convection via CAPE, conditions do not produce an LCL–ABL crossing due to the elevated LCL, preventing feedback. These results imply that even when CAPE is very large, tropospheric atmospheric conditions may still suppress deep convection. This comparison may provide a quantitative perspective on feedback mechanisms across different environments; more importantly, these results highlight the dual importance of both surface fluxes and atmospheric properties in convective triggering on climatologically relevant timescales.

\section{Conclusion}

This study presents a parsimonious framework that links soil moisture, vegetation hydraulics, atmospheric boundary layer (ABL) dynamics, convective available potential energy (CAPE), and stochastic rainfall, building on previous work \citep{yin2015land, emanuel2023physics}. A central result is that plant hydraulic traits and soil hydraulic properties do not simply influence convection through mean evapotranspiration, but more dynamically through their control of soil moisture decline, surface-flux partitioning, boundary-layer growth, and the timing of ABL--LCL interaction. In this way, the land surface regulates both the likelihood of convective initiation and the thermodynamic intensity of the resulting event. The diurnal simulations (Figure \ref{fig:sidebyside}) show that, under the same atmospheric forcing, differences in soil moisture and plant water-use strategy alter stomatal conductance, shift the relative evolution of ABL height and LCL height, and thereby change the CAPE available at the time of crossing. The dry-down analysis (Figures \ref{fig:realdrydown} and \ref{fig:soiltype}) further shows that these effects organize into systematic soil- and plant-dependent CAPE--soil moisture relationships, providing an emergent link between ecohydrological state and convective potential.

Our results show that soil properties strongly modulate the evolution of convective potential during drying. Coarser and intermediate soils can sustain relatively large CAPE under wetter conditions, but differ markedly in how quickly this potential declines as the root zone dries. In particular, loamy sand maintains high CAPE over a broader portion of the dry-down, whereas sandy soils lose convective potential more rapidly because plant-available water is depleted faster; in finer soils, increasingly negative matric potentials suppress transpiration and convection as drought intensifies. Plant hydraulic behavior adds a second axis of regulation: traits associated with high maximum stomatal conductance promote strong humidification of the boundary layer and high CAPE under well-watered conditions, whereas more conservative or drought-tolerant strategies prolong transpiration and maintain convective potential deeper into the dry-down. Thus, convection is controlled not by soil moisture alone but by how soil and plant traits together shape water potential gradients, latent and sensible heat fluxes, ABL growth, and CAPE buildup.

By embedding this land--atmosphere coupling into a stochastic rainfall framework, we show that the same mechanisms operating on the diurnal scale can feed back onto seasonal and multi-year soil moisture dynamics. Because both rainfall frequency and intensity depend on the CAPE generated at the land surface, the model produces persistent wet and dry regimes, along with switching between them. The long-term simulations (Figures \ref{fig: rainfallfreqgraph} and \ref{fig: rainfallfreqgraphlatitude}) show that vegetation type, rooting depth, hydraulic traits, and background atmospheric conditions can substantially alter the persistence of these regimes. In particular, deeper-rooted and hydraulically conservative vegetation is more likely to sustain wet persistence by maintaining sufficient transpiration and convective triggering over longer periods, whereas shallow-rooted or more aggressively water-using vegetation can favor faster drying and shorter-lived wet states. These results provide a mechanistic explanation for how ecohydrological traits may contribute to hydroclimatic persistence.

More broadly, we developed a process-based stochastic model in which convective rainfall is interpreted as an emergent outcome of coupled soil--plant--boundary-layer dynamics, superimposed on an external precipitation forcing independent of land-surface state. This perspective helps connect plant function, soil hydrology, and atmospheric convection within a common dynamical framework, and clarifies how land-surface characteristics can influence both short-term convective initiation and long-term climate variability.

Several extensions of this framework remain important. First, the present analysis isolates the temporal dimension of land--atmosphere feedback under horizontally homogeneous conditions; future work should examine how spatial heterogeneity in soils, vegetation, and topography perturbs the baseline mechanisms identified here \cite{hohenegger2018role}. Soil hydrologic dynamics can be realistically applied at the regional watershed scale, where averaged surface properties capture the aggregate effects of local feedbacks \citep{bartlett2025stochastic}. Second, while the free atmosphere is treated as prescribed to isolate land-surface effects, allowing free-atmospheric humidity and stability to evolve in time would provide a more complete closure of the coupled water and energy budgets. Third, additional atmospheric constraints on convective initiation, including CIN, mesoscale organization, and horizontal advection, should be incorporated to better assess when high CAPE generated by the land surface does or does not translate into rainfall. Even with these limitations, the framework developed here highlights the important role of plant and soil traits in shaping convective rainfall and suggests that shifts in vegetation composition, rooting strategies, and soil hydraulic conditions may have important consequences for hydroclimatic persistence under environmental change.

\section*{\textit{Acknowledgments}}
This work was funded through a cooperative endeavor agreement between The Water Institute and the Louisiana Office of Community Development as part of the Louisiana Watershed Initiative (LWI) funded via Catalog of Federal Domestic Assistance (CFDA) 14.228 Grant B-18-DP-22-0001. We also thank the Carbon Mitigation Initiative for funding this project. 

\section*{\textit{Data Availability}}
Upon final revision, code for model simulation will be deposited in GitHub and Zotero in accordance with FAIR principles. The linear fit in Eq.~\eqref{eq: closeeq} was derived from measurements and plots reported in \cite{chen2019prediction}. The data used in Figure \ref{fig: stomcurve} was obtained from paper \cite{henry2019stomatal} with the original source cited therein. The data used to derive the free atmospheric parameters, from \cite{cerasoli2021cloud}, is available at: [\url{https://www.ecmwf.int/en/forecasts/dataset/coupled-reanalysis-20th-century}]. Early-morning atmospheric parameters in the dry-down section are from Radiosonde data from the ARM program [\url{http://www.arm.gov/}], accessed through \cite{yin2015land}.

\appendix
\section{Precipitation Intensity Thermodynamics}
\label{app:PI APP}
The derivation for convective precipitation totals, adapted from \cite{muller2011intensification}, will begin with the first law of thermodynamics for a moist pseudo-adiabatic process,
\begin{equation}
c_p\, dT - \alpha_v\, dp = - L_v\, dq_{\mathrm{sat}},
\label{eq: first}
\end{equation}
where $q_{\text{sat}}$ is the saturation specific humidity, $\alpha_v$ is the specific volume, as the specific humidity and water vapor mixing ratio are approximately equal ($q \approx w$) in typical atmospheric conditions. The latent heat of vaporization is denoted by $L_v$.

With the hydrostatic assumption along with Eq.~\eqref{eq: first}, the derivative of moist static energy ($MSE$) with respect to $z$ can be defined as 
\begin{equation}
\label{eq: MSE1}
\frac{dMSE}{dz} = L_v\frac{dq_{sat}}{dz}+g+ c_p\frac{dT}{dz} = 0.
\end{equation}

Given that moist static energy is conserved during pseudo-adiabatic ascent, we obtain the vertical dry static energy relation from Eq.~\eqref{eq: MSE1}. If we define the dry static energy as $S_d = gz + c_pT$, we find the relation: 

\begin{equation}
\label{eq: MSE}
\frac{dMSE}{dz} = L_v\frac{dq_{sat}}{dz}+\frac{dS_d}{dz} = 0.
\end{equation}

Separately, we may define the vertical dry static budget \citep{muller2011intensification}:
\begin{equation}
\label{eq: vsb}
\underbrace{
\int_{z^*_{\text{LCL}}} ^{z^*_{\text{LNB}}}   
\rho_a(z)\left(\frac{\partial S_d}{\partial t}+w(z)\, \frac{\partial S_d}{\partial z}\right)  dz
}_{\text{condensation, $C_n$}}
=
\underbrace{
L_v \int_{z^*_{\text{LCL}}} ^{z^*_{\text{LNB}}}\rho_a(z)\left(\frac{\partial \sum q_n}{\partial t}+w(z)\, \frac{\partial \sum q_n}{\partial z} \right)dz
}_{\text{storage tendency, $S$}}
+ L_v I_c .
\end{equation}

where $w(z)$ is the updraft velocity in the troposphere, $L_v$ is the latent heat of vaporization, and $\sum q_n$ are the respective mass fractions of non-precipitating water (e.g., snow, rain, ice, and graupel), assuming the latent heats of sublimation and vaporization are approximately equal. The square brackets express the vertically weighted mass integral. If $\frac{dS_d}{dz}$ in Eq.~\eqref{eq: vsb} is replaced with $-L_v\frac{dq_{sat}}{dz}$ from Eq.~\eqref{eq: MSE} and we assume the local time rate of change of $S_d$ is approximately zero, we may then solve for $I_c$, the convective precipitation intensity. As in \cite{muller2011intensification}, we define a storm static efficiency as a function of net storage changes (i.e., the first bracketed term on the right-hand side of Eq.~\eqref{eq: vsb}), and graupel, given as $S$. Net condensation, i.e. $-\int_{z^*_{\text{LCL}}} ^{z^*_{\text{LNB}}}\left(w(z)\frac{\partial q_{\text{sat}}}{\partial z } \right)\rho_a(z)dz$, is denoted as $C_n$:
\begin{equation}
\label{eq: effic}
\varepsilon' = 1 - \frac{S}{C_n},
\end{equation}
where the intensity is defined as 
\begin{equation}
\label{eq: efficIC}
-\varepsilon'\sqrt{2\eta}\int_{z^*_{\text{LCL}}} ^{z^*_{\text{LNB}}}\left(w(z)\frac{\partial q_{\text{sat}}}{\partial z } \right)\rho_a(z)dz = I_c,
\end{equation}
found with algebraic manipulation. A higher efficiency means that more water immediately falls to the surface as rain rather than being stored or advected from the atmosphere. 

The dynamic nature of this efficiency term is important for land--atmosphere interactions. However, for the purposes of this paper, we consider a constant ratio of condensation to evaporation and sublimation.

\section{Lifting Condensation Level Height and Atmospheric Variables}
\label{app: LCLh}

The LCL height, derived analytically from \cite{yin2015land}, is given as a function of the near-surface temperature and specific humidities, 

\begin{equation}
z_{L C L}=-\frac{T_0}{\Gamma_{\text{dry}}}+\frac{\lambda R}{g R_v W\left[\frac{\lambda R \Gamma_{\text{dry}}}{g R_v T_0}\left(\frac{P_0 q_0}{\varepsilon e_{\text{ref}}}\right)^{-\frac{R \Gamma_{\text{dry}}}{g}} \exp \left(\frac{\lambda R \Gamma_{\text{dry}}}{g R_v T_{ \text{ref}}}\right)\right]},
\label{eq: lclh}
\end{equation}

given as $T_0$ and $q_0$, respectively. These can be replaced with the corresponding $\theta_{\text{BL}}$ and $q_{\text{BL}}$ values.

\begin{figure}[h!]
        \centering
        \includegraphics[width=.7\linewidth]{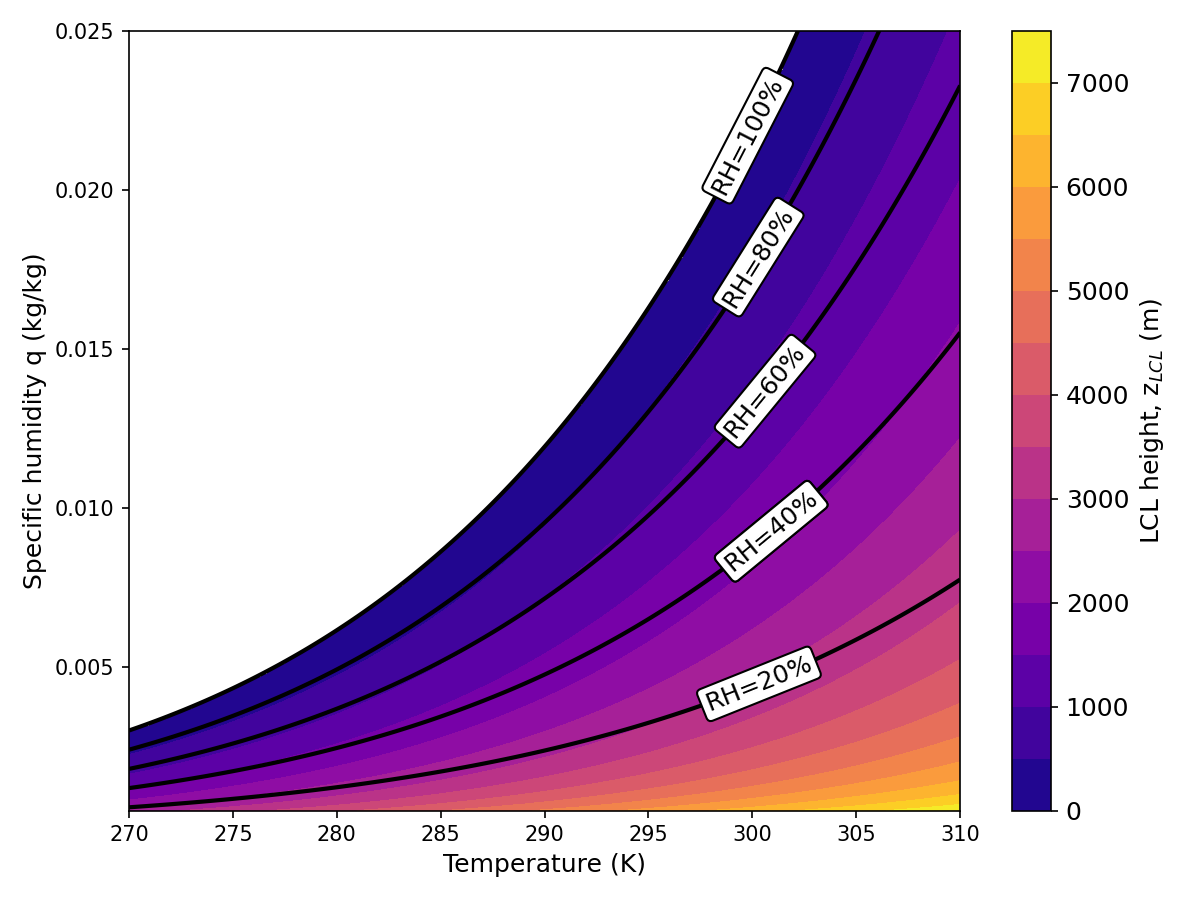}
        \caption{Visualization of $z_{\text{LCL}}$ given from the formulation of Eq.~\eqref{eq: lclh}, dependent on surface temperature and humidity (Eqs.~\eqref{eq: dqdt} and~\eqref{eq: dthetadt}).}.
        \label{fig: diagram LCL}
\end{figure}

The saturated lapse rate for the parcel is given as 

\begin{equation}
\frac{d T_p}{dz}=-\frac{g}{c_p} \frac{1+\frac{q_{\text{sat}}}{RT}}{1+\frac{L_v^2 q_{\text{sat}}}{c_p R_v T^2}},
\label{eq: moist lapse rate}
\end{equation}

while the free atmospheric virtual potential temperature and specific humidity profiles are assumed to have a linear form of

\begin{equation}
T_{f}=\theta_{v,f,0}+\gamma_{\theta_{v,f}}z,
\label{eq: free lapse rate}
\end{equation}

and similarly 

\begin{equation}
q_{f}=q_{0,f}+\gamma_{q,f}z
\label{eq: free lapse rate humid}.
\end{equation}

\section{Soil-Plant Model}
\label{app: spmodel}
In Eq.~\eqref{equationtrans}, the volumetric liquid flux is multiplied by the conductance $g_{srp} = \frac{LAIg_pg_{sr}}{LAIg_p+g_{sr}}$. The soil water potential is linked to $s$ by the so-called retention curve \begin{equation}
\psi_s(s)=\overline{\psi_s} s^{-b},
\end{equation} with parameters $\overline{\psi_s}$ and $b$ depending on soil type.

Given Eq.~\eqref{balance-eq}, we assume that all excess water above soil saturation is lost as runoff, represented by $Q$. In highly vegetated regions, Hortonian runoff, otherwise known as infiltration-excess runoff, is assumed to be negligible \citep{porporato2022ecohydrology}. Leakage is equal to the soil hydraulic conductivity, $K=K_s s^{2 b+3}$, with the curve also dependent on soil type.
The soil-root conductance is given by the following formulation \citep{katul2003relationship, daly2004coupled, bartlett2014coupled},

\begin{equation}
g_{s r}=\frac{K \sqrt{RAI_w s^{-a}}}{\pi g \rho_w Z_r},
\end{equation}

where $K$ is the hydraulic conductivity, $RAI_w$ is the root area index under well-watered conditions, and $a$ is a species-dependent parameter. The plant conductance \citep{daly2004coupled, hartzell2017role} also depends on the stem vulnerability to cavitation, and is given as 
\begin{equation}
g_p=g_{p, \max }\left[1+\left(\frac{\psi_{\text{stem}}}{\psi_{50,\text{stem}}}\right)^c\right]^{-1}.
\end{equation}

Here, $c$ is taken to be 4, as in \cite{hartzell2017role}. The model for stomatal conductance is given as in \cite{jarvis1976interpretation, daly2004coupled}, 
\begin{equation}
g_{st}=g_{st, \max } f_{\phi}(\phi) f_{\theta_{BL}}\left(\theta_{B L}\right) f_{\psi_l}\left(\psi_l\right) f_{D}(D),
\label{eq: jarv}
\end{equation}

where the slope and intercept of $f_{\psi_l}$ is found from Eq.~\eqref{eq: closeeq}, given the stem 50$\%$ loss of hydraulic conductivity mark. The variable $D$ is the vapor pressure deficit, i.e., $e_{\text{sat}} - e_{\text{BL}}$. 

\begin{figure}[h]
\centering
\includegraphics[width=0.6\textwidth]{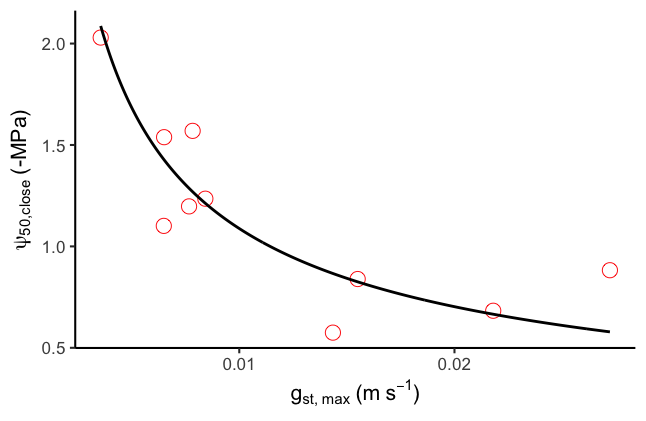}
\caption{Coordination of plant physiological traits, input into Eq.~\eqref{eq: jarv}. Blue lines describe the fitted relationship between maximum stomatal conductance and $\psi_{50,\mathrm{stem}}$ taken from data \citep{hao2010differentiation}, using the R package SMATR as in \cite{henry2019stomatal}.}
\label{fig: stomcurve}
\end{figure}

The formulation for the atmospheric conductance is 

\begin{equation}
g_{\mathrm{a}}= \frac{\bar{u}_a k^2}{\left(\ln \left[\frac{z_a-d}{\epsilon}\right] \ln \left[\frac{z_a-d}{\epsilon_q}\right]\right)}
\end{equation}

where $u_a$ is the wind velocity at a reference level $z_a$, $k$ = 0.4 is the Von Karman constant $\epsilon_q$ is the water vapor roughness height, $\epsilon$ is the momentum roughness height, and $d$ is the displacement height.

\bibliographystyle{ametsocV6}
\bibliography{referenceslibby}

\end{document}